\documentclass[prd,twocolumn,nopacs,floatfix,amsmath,nofootinbib,amssymb,preprintnumbers]{revtex4-2}
\usepackage{subcaption}
\usepackage{color,dcolumn,booktabs,diagbox}
\usepackage[final]{graphicx}
\usepackage{pgfplots}
\pgfplotsset{compat=newest}
\usepackage{txfonts}
\usepackage{slashed}
\usepackage{epstopdf}
\usepackage{multirow}
\usepackage{bm}
\usepackage{pifont}
\usepackage{adjustbox}
\usepackage{siunitx}
\usepackage{multirow}
\usepackage{makecell}
\usepackage{rotating}
\usepackage{verbatim}

\usepackage{graphicx,xcolor,dcolumn,booktabs}
\usepackage[colorlinks,
            citecolor=blue,
            anchorcolor=red,
            menucolor=red,
            linkcolor=red,
            filecolor=red,
            runcolor=red,
            urlcolor=blue,
            frenchlinks=true]{hyperref}
\begin{document}

\title{Assessing the validity of the Born-Oppenheimer approximation in\\ potential models for doubly heavy hadrons}

\author{Zi-Long Man$^{1,2,3,4}$}
\author{Hao Zhou$^{1,2,3,4}$}
\author{Si-Qiang Luo$^{1,2,3,4}$}
\author{Xiang Liu$^{1,2,3,4}$}
\email{xiangliu@lzu.edu.cn}

\affiliation{
$^1$School of Physical Science and Technology, Lanzhou University, Lanzhou 730000, China\\
$^2$Lanzhou Center for Theoretical Physics,
Key Laboratory of Theoretical Physics of Gansu Province,
Key Laboratory of Quantum Theory and Applications of MoE,
Gansu Provincial Research Center for Basic Disciplines of Quantum Physics, Lanzhou University, Lanzhou 730000, China\\
$^3$MoE Frontiers Science Center for Rare Isotopes, Lanzhou University, Lanzhou 730000, China\\
$^4$Research Center for Hadron and CSR Physics, Lanzhou University and Institute of Modern Physics of CAS, Lanzhou 730000, China}

\begin{abstract}
The Born–Oppenheimer approximation is widely used to investigate the properties of hydrogen-like systems and doubly heavy hadrons. However, the extent to which this approximation captures the features of such systems within potential models remains an open question. In this work, we adopt the results obtained with the Gaussian expansion method as a benchmark to assess the validity of the Born–Oppenheimer approximation within potential models for hadronic systems. We also investigate the dependence of the Born-Oppenheimer approximation  results on the choice of trial wave functions. A comprehensive study of the Born-Oppenheimer approximation is carried out by performing  calculations using Slater-type functions and Gaussian-type functions as trial wave functions, and by comparing the resulting predictions with those obtained from the Gaussian expansion method. We find that the calculations performed within the Born-Oppenheimer approximation are close to those obtained with the Gaussian expansion method when the heavy-quark mass is relatively small. However, as the heavy-quark mass increases, calculations employing Slater-type functions yield larger values than those from the Gaussian expansion method, whereas those using Gaussian-type functions lead to smaller ones. The use of Slater-type functions generally leads to an enhanced binding energy. The underestimation observed in Born-Oppenheimer approximation calculations with Gaussian-type functions  primarily stems from the neglect of non-adiabatic corrections. This comparative study provides deeper insight into the structure of doubly heavy hadrons and helps clarify the applicability and limitations of the Born–Oppenheimer treatment within potential models.
\end{abstract}

\maketitle
\newcommand{\bfxA}{\mathbf{x_A}}
\newcommand{\bfxB}{\mathbf{x_B}}
\newcommand{\bfxone}{\mathbf{x_1}} 
\newcommand{\bfxtwo}{\mathbf{x_2}} 

\section{Introduction}

At present, we are entering an era of high-precision hadron spectroscopy. This is evidenced not only by advancements in experimental precision—with the observation of $\Xi_{cc}^{++}(3621)$ serving as a prime example discussed below—but also by continuous improvements in theoretical calculational accuracy. In this work, we focus on examining the applicability of the Born–Oppenheimer approximation (BOA) to doubly  heavy hadronic systems.

In 2002, the SELEX Collaboration reported the observation of a doubly charm baryon, $\Xi_{cc}^+(3519)$, in the $\Lambda_{c}^+K^-\pi^+$ final state~\cite{SELEX:2002wqn}. Its mass was measured as $3519 \pm 2$~MeV, with a lifetime constrained to $< 33$~fs at the 90\% confidence level. This observation was later corroborated by SELEX in an analysis of the $pD^+K^-$ decay channel. However, subsequent experiments by other collaborations did not confirm the existence of $\Xi_{cc}^+(3519)$ as reported by SELEX~\cite{Ratti:2003ez,BaBar:2006bab,Belle:2006edu}, highlighting the need for higher-precision data to resolve this discrepancy.

With the accumulation of experimental data, in 2017, the LHCb Collaboration investigated the $\Lambda^+_cK^-\pi^+\pi^+$ process and reported the observation of the $\Xi_{cc}^{++}(3612)$ baryon, with a mass and lifetime measured to be $3621.55\pm0.23\pm0.30$~MeV and $0.256^{+0.024}_{-0.022}\pm0.014$~ps, respectively~\cite{LHCb:2019epo,LHCb:2018zpl}. This resonance has since been confirmed in multiple decay modes, including $\Xi_c^+\pi^+$~\cite{LHCb:2018pcs}, $\Xi_c^{\prime+}\pi^+$~\cite{LHCb:2022rpd}, and $\Xi_c^0\pi^+\pi^+$~\cite{LHCb:2025shu}. The confirmed existence of a doubly  charm baryon demonstrates that two identical heavy quarks can be bound inside a hadron, strengthening the motivation to search for doubly  heavy tetraquark states. In 2022, the LHCb Collaboration reported the first observation of a doubly  charm tetraquark, the $T_{cc}^{+}(3875)$, in the $D^0D^0\pi^+$ mass spectrum~\cite{LHCb:2021vvq,LHCb:2021auc}. In recent years, the discoveries of doubly  charm baryons and tetraquarks have greatly expanded the scope of hadron spectroscopy and stimulated extensive discussions in both theory and experiment~\cite{Chen:2016qju,Liu:2019zoy,Chen:2022asf,Liu:2024uxn,Wang:2025dur}.

In the past years, the BOA~\cite{Born:1927rpw} has been widely used in the study of
doubly heavy hadrons. 
On the one hand,
this methodology has been rigorously formulated and extensively studied within QCD-based approaches in Refs.~\cite{Juge:1999ie,Braaten:2014qka,Bicudo:2015vta,Bicudo:2016ooe,Bruschini:2023zkb,Berwein:2024ztx,Braaten:2024stn,Braaten:2024tbm}.
On the other hand, the BOA has also been implemented within potential models (see, e.g., Refs..~\cite{Maiani:2019cwl,Grinstein:2024rcu,Liu:2025jyn,Germani:2025mos,Kang:2025xqm} ), where effective potentials are employed to describe the dynamics of doubly heavy hadrons.
In Refs.~\cite{Maiani:2019cwl,Liu:2025jyn}, for instance, the BOA was applied to study doubly  heavy baryons and tetraquarks.
 The mass of the $\Xi_{cc}$ obtained in Ref.~\cite{Maiani:2019cwl} agrees with the measured value of $\Xi_{cc}^{++}(3612)$, and Ref.~\cite{Liu:2025jyn} further suggests that the observed $T_{cc}^{+}(3875)$ is a good candidate for the $cc\bar{n}\bar{n}$ tetraquark state with quantum numbers $(I,J^P)=(0,1^+)$. However, in both works, the confining interaction between the heavy and light quarks was not fully incorporated. Since confinement plays a crucial role in determining heavy–light separations, omitting part of this interaction introduces non-negligible uncertainties into the predicted masses. Moreover, their results rely on an extra adjustable parameter $R_0$ to constrain the estimated masses. 
A complete treatment of the heavy–light confining interaction, without introducing additional parameters, is therefore essential for obtaining reliable predictions for doubly  heavy systems.

The BOA was originally developed in nonrelativistic quantum mechanics to describe molecules, where light electrons move in the field of much heavier nuclei. By effectively separating the motions of the light and heavy degrees of freedom, the two parts can be treated independently and solved sequentially. This approach provides an accurate description of systems such as the hydrogen molecule and the hydrogen molecular ion~\cite{pauling1928application}. Its success relies on the fact that the proton mass is approximately 1800 times larger than the electron mass. In contrast, for doubly  charm hadronic states the situation is markedly different: the constituent mass of the $c$ quark is only about five times that of the $n$ (i.e., $u$ or $d$) quark. This significant difference naturally raises the question of whether the BOA can be reliably applied to determine the properties of doubly  heavy hadrons, an issue that requires careful examination. 
Moreover, the selection of basis functions within the BOA framework can lead to discernible variations in the numerical results. In molecular physics, Slater-type functions (STFs) are generally preferred over Gaussian-type functions (GTFs), due to their superior capability in reproducing the finite slope behavior of wave functions at short distances, and they also represent the exact solutions to the hydrogen atom problem \cite{szabo2012modern}. Nevertheless, the applicability and efficiency of STFs within the realm of hadron physics, where Gaussian type functions are conventionally favored for their rapid long-range suppression. This potential discrepancy warrants a systematic comparative study.

Compared with the BOA, the Gaussian expansion method (GEM) provides a fully dynamical treatment by solving the Schr\"{o}dinger equation without assuming a separation of the heavy and light mass degrees of freedom~\cite{Hiyama:2003cu}. The GEM delivers high numerical precision for few-body systems and incorporates the full correlations among the constituent quarks. Therefore, a direct comparison between the BOA and the GEM offers an effective way to quantify the uncertainties inherent in the BOA and to clarify the conditions under which it remains valid for doubly  heavy systems. Such a comparison is essential for establishing reliable theoretical predictions for doubly  heavy baryons and tetraquarks.

In this paper, we first investigate the hydrogen molecular ion and hydrogen molecule using both the BOA and the GEM. Within the BOA framework, STFs and GTFs were also employed as trial wave functions, and the corresponding results are compared with those obtained from the GEM.
We demonstrate that the accuracy of the BOA depends sensitively on the mass hierarchy between the heavy and light constituents: a pronounced scale separation yields results close to the solutions of GEM, while  deviations arise as this hierarchy becomes less distinct. 

This study focuses on the validity of the BOA within the framework of potential models, rather than the QCD-based studies mentioned above. Therefore, we present a general Hamiltonian applicable to both doubly heavy baryons and tetraquarks, and investigate these systems using both the BOA and the GEM. In hadronic systems, our results indicate that the three approaches yield close values when the heavy-quark mass is relatively small. However, as the heavy-quark mass increases, opposite trends emerge for the two basis choices: the BOA-STFs calculations produce larger values than those obtained from the GEM, whereas the BOA-GTFs calculations yield smaller ones.


The remainder of this paper is organized as follows. In Sec.~\ref{secII}, a detailed comparison of the BOA-STFs, BOA-GTFs, and GEM results for hydrogen-like systems is presented.
Sec.~\ref{secIII} introduces the  Hamiltonian for doubly  heavy baryon and tetraquark. Based on the this Hamiltonian, we describe the numerical approaches used in this work. Three methods are applied to estimate the masses of doubly  heavy baryons and tetraquark states. A summary of this work is given in Sec.~\ref{secV}.
Technical details are expanded in the appendix.
\section{hydrogen molecular ion and hydrogen molecule} \label{secII}

\begin{figure}[!htbp]
	\centering
	\begin{minipage}[b]{0.48\textwidth}
		\centering
		\includegraphics[width=1\textwidth]{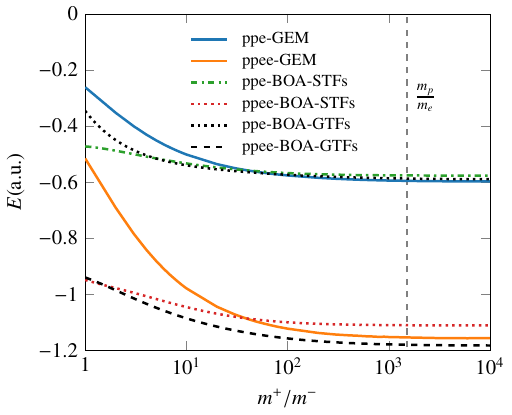}
	\end{minipage}
 \caption{The eigenvalues of the hydrogen molecular ion and hydrogen molecule as function of the mass ratio $m_i/m_e$, obtained within the BOA-STFs, BOA-GTFs, and GEM. The vertical dashed line denotes the physical value $m_p/m_e$.}\label{ppe}
\end{figure}

In order to obtain more reliable mass estimates, we solve the multi-body Schr\"{o}dinger equations using the Hamiltonian  within the  GEM~\cite{Hiyama:2003cu,Hiyama:2012sma}. This method provides high numerical accuracy and has been widely applied in hadron masses spectroscopy \cite{Luo:2023sne,Man:2024mvl,Zhou:2025fpp,Luo:2025psq,Zhang:2025vtc,Luo:2025cqs,An:2025rjv,Meng:2023jqk,Wu:2024hrv,Wu:2024euj,Wu:2024zbx}.
In recent years, we have successfully developed  theoretical descriptions of the mass spectra of singly heavy baryons \cite{Luo:2023sra,Luo:2023sne,Peng:2024pyl,Luo:2025cqs}, 
$\Omega$ hyperons, and triply heavy baryons using the GEM \cite{Zhou:2025fpp}. In addition,  a new charm-strange baryon state was predicted~\cite{Luo:2021dvj}. Furthermore,
 the GEM has been employed to study the spectra of the hydrogen atom, molecular ion, and diatomic molecule, with results found to be in excellent agreement with experimental measurements~\cite{Luo:2023hnp}. We have also predict a new type of hydrogen-like charm-pion or charm-kaon matter. 
These studies demonstrate that the GEM adopted in our framework provides a reliable and accurate description of mass spectra.

For hydrogen-like systems, GEM reproduces experimental binding energies with an accuracy better than BOA, which validates its role as a numerical benchmark in the present comparative study.
In  molecular physics,  the Hamiltonians of hydrogen molecular ion and hydrogen molecule can be written as 
\begin{equation}\label{totH}
	H=\sum_i\frac{p_i^2}{2m_i/m_{e}}+\sum_{i<j}\frac{1}{r_{ij}},
\end{equation}
where $m_i$ and $m_{e}$ denotes the masses of the $i$-th  proton and electron, respectively.

The comparison is shown in Fig.~\ref{ppe}. 
It is  found that when the mass ratio between the heavy and light particles is large, the results obtained from both BOA-STFs and BOA-GTFs approach those of GEM. However, as the mass ratio decreases, the BOA predictions increasingly deviate from the GEM results and tend to yield deeper binding energies for both hydrogen molecular ion and  hydrogen molecule.

The main reason for the discrepancy between the results of the BOA and the GEM originates from the limited validity of the BOA. 
According to Ref.~\cite{Weinberg_2015},  the applicability of the BOA  to molecular systems is governed by 
\begin{equation}\label{ratio}
	\left(\frac{m_e}{m_i}\right)^{1/4}.
\end{equation}
When $m_e/m_i \sim 1$, the underlying assumption of scale separation breaks down and the BOA becomes unreliable. The BOA approach generally provides a lower estimated mass \cite{Bratsev:1965,Das:1993zz}.
In contrast, for $m_e/m_i \ll 1$, the BOA is expected to be valid and its predictions approach the GEM solutions.

The above analysis of the hydrogen molecular ion and the hydrogen molecule provides a benchmark for assessing the validity of the BOA. It explicitly demonstrates that the reliability of the BOA is controlled by the degree of mass hierarchy between the heavy and light constituents: when a clear separation of scales exists, the BOA yields results close to the GEM solutions, whereas significant deviations emerge once this hierarchy is weakened.

This observation is directly relevant to doubly  heavy systems in quantum chromodynamics. In doubly  heavy hadrons, such as doubly  heavy baryons and tetraquarks, the heavy quarks play a role analogous to the nuclei in molecular systems, while the light quarks correspond to the electrons. However, unlike molecular physics, the mass ratio between heavy and light quarks is much smaller.  Moreover, strong interaction contains an intrinsic color-confining property that has no counterpart in atomic systems. These features raise nontrivial questions regarding the quantitative reliability of the BOA when applied to doubly  heavy hadrons.

\section{doubly  heavy baryons and tetraquarks } \label{secIII}

\subsection{General Hamiltonian}

In order to study the  masses of $S$-wave doubly  heavy hadronic states, a general Hamiltonian is given by
\begin{equation}\label{totH}
	H=\sum_i\frac{p_i^2}{2m_i}+\sum_{i<j}(V_{ij}^{\text{conf}}+ V_{ij}^{\text{hyp}}),
\end{equation}
where the subscript $i$ represents the $i$-th constituent quark. $p_i$ is its constituent quark  momentum. Since we focus on  $S$-wave doubly  heavy hadrons, the confinement term $V_{ij}^{\text{conf}}$ and hyperfine term $ V_{ij}^{\text{hyp}}$ take the following forms:

\begin{equation}
		V_{ij}^{\text{conf}}=-\left[-\frac{\alpha_{s}}{r_{ij}}+\frac{3}{4}br_{ij}+\frac{3}{4}C\right]\boldsymbol{F}_{i}\boldsymbol{\cdot}\boldsymbol{F}_{j},
	\end{equation}
 \begin{equation}
     V_{ij}^{\text{hyp}}=\left[\frac{8\pi\alpha_{s}}{3m_{i}m_{j}}\tilde{\delta}(r_{ij})\boldsymbol{S}_{i}\boldsymbol{\cdot}\boldsymbol{S}_{j}\right]\boldsymbol{F}_{i}\boldsymbol{\cdot}\boldsymbol{F}_{j},
 \end{equation}
respectively. The $\alpha_s$, $b$, and $C$ are the strong coupling constant, the string tension, and the renormalization constant, respectively.
The smeared delta function $\tilde{\delta}(r_{ij})$ is defined as
\begin{equation}
	\tilde{\delta}(r_{ij})=\frac{\sigma^3}{\pi^{3/2}}e^{-\sigma^2r^2_{ij}}.
\end{equation}
Here, $\sigma$ is smearing parameter.

\subsection{Doubly  heavy systems within BOA} 

According to Refs. \cite{Maiani:2019lpu,Maiani:2019cwl,Liu:2025jyn},  within BOA the Hamiltonian can be decomposed into heavy and light components,
\begin{equation}\label{BOpotential}
	H_{\text{BOA}}=H_{heavy}+H_{light}.
\end{equation}
The interaction of heavy degree of freedom is given by
\begin{equation}
	H_{heavy}= T_{heavy}+V(r_{AB}),
\end{equation}
where the heavy–heavy interaction potential takes the form
\begin{equation}
	V(r_{AB})= -\frac{2\alpha_s}{3 r_{AB}}+ \frac12br_{AB}+\frac12 C.
\end{equation}
The interaction associated with the light degree of freedom can be written as
\begin{equation}
	H_{light}= T_{light}+V_l.
\end{equation}

For the doubly heavy baryon states, the interaction for light degree of freedom is given by 
\begin{equation}
	V_l=-\frac{2\alpha_s}{3r_A}-\frac{2\alpha_s}{3r_B}+\frac12 br_A+\frac12 br_B+C.
\end{equation}

The corresponding Schr\"{o}dinger equation reads as
\begin{equation}\label{totSh}
\left(T_{heavy}+T_{light}+V(r_{AB})+V_l(r_A,r_B,r_{AB}) \right)\Psi=E\Psi,
\end{equation}
with the total wave function written as
\begin{equation}\label{totwave}
 \Psi=f(r_{AB})\Phi(r_A,r_B,r_{AB}).
\end{equation}
Here, $f(r_{AB})$ and $\Phi(r_A,r_B,r_{AB})$ denote the wave functions associated with the heavy and light degrees of freedom, respectively.

The BOA relies on the suppression of the  kinetic energy of the heavy degrees of freedom due to their large masses. Therefore, we first  consider only the light part by solving the eigenvalue equation of the light particles for the fixed positions of the heavy particles
\begin{equation}\label{lightSh}
\left(T_{light}+V_l \right)\Phi=\varepsilon(r_{AB}) \Phi,
\end{equation}
where $\varepsilon(r_{AB})$ denotes the energy eigenvalue of the light degrees of freedom for a fixed heavy–quark separation.
With $\varepsilon(r_{AB})$  determined, Eq.~(\ref{totSh})  reduces to Schr\"{o}dinger equation for the heavy degree of freedom,
\begin{equation}
\left(\frac{-\nabla^2}{2m_Q}+V(r_{AB})+\varepsilon(r_{AB})\right)f(r_{AB})=E_{BOA} f(r_{AB}).
\end{equation}

Finally, the resulting BOA Hamiltonian of doubly heavy baryon is 
\begin{equation}
	H_{BOA}= -\frac{\nabla^2}{2m_Q}-\frac{2\alpha_s}{3r_{AB}} +\frac12br_{AB}+\frac12 C+\varepsilon(r_{AB}).
\end{equation}
With the BOA Hamiltonian, we can solve the Schr\"{o}dinger
 equation to obtain the eigenvalue of the doubly heavy baryon: 
\begin{equation}
	H_{BOA}f(r_{AB})=E_{BOA}f(r_{AB}),
\end{equation}
where $E_{\text{BOA}}$ is the BOA eigenvalue of the discussional system.

For the doubly heavy tetraquarks, the $QQ$ diquark can reside in either the color $\bar{3}$ or 6 representation. In present work, we firstly focus on the  color $\bar{3}$ representation and the heavy-heavy quark  potential takes the form
\begin{equation}
V(r_{AB})=-\frac{2\alpha_s}{3 r_{AB}}+ \frac12br_{AB}+\frac12 C.
\end{equation}

The interaction associated with the light degrees of freedom is given by
\begin{equation}
\begin{split}
	V_l&=
 -\frac{1\alpha_s}{3r_{A1}}-\frac{1\alpha_s}{3r_{B2}}
  -\frac{1\alpha_s}{3r_{A2}}-\frac{1\alpha_s}{3r_{B1}}-\frac{2\alpha_s}{3r_{l}}\\
  &\quad+\frac14 b r_{A1}+\frac14 b r_{B2}+\frac14 b r_{A2}+\frac14 b r_{B1}+C.
  \end{split}
\end{equation}
Here, $r_{A1}$, $r_{A2}$, $r_{B1}$, $r_{B2}$ represent the distances  between the heavy quarks and the light quarks.
$r_{l}$  is the distance between two light quarks. The eigenvalue of the light degree of freedom $\varepsilon(r_{AB})$ for the doubly heavy tetraquark is obtained using Eq.~(\ref{lightSh}).

Within the BOA framework, the resulting BOA Hamiltonian for the doubly heavy tetraquark can be written as
\begin{equation}
	H_{BOA}= -\frac{\nabla^2}{2m_Q}+V(r_{AB})+\varepsilon(r_{AB}).
\end{equation}
The selection of the wave function and its detailed derivation are presented in the Appendix.

Following  Refs. \cite{Maiani:2019lpu,Liu:2025jyn}, the hyperfine structure of doubly heavy system is described within the framework of chromomagnetic interaction (CMI) model~\cite{Liu:2019zoy,Li:2023wug,Li:2023wxm,Li:2023aui,Li:2025fmf}.
Adding hyperfine interactions, the mass formula of doubly heavy system is given by
\begin{equation} \label{mass_baryon}
	M=\sum_{i}m_i+E_{BOA}+\langle H_{CMI}\rangle.
\end{equation}
  $H_{CMI}$ represents color-magnetic interactions (CMI) between quark components, which has the form 
\begin{equation}\label{CMI}
	H_{CMI}=-\sum_{i<j}C_{ij}\lambda_i\cdot\lambda_j\sigma_i\cdot\sigma_j.
\end{equation}
Here, $\lambda_i$ and $\sigma_i$ denote the $SU(3)$ Gell-Mann matrices and SU(2) Pauli matrices for the $i$th quark, respectively. The coupling constant $C_{ij}$ describes the strength between the $i$th quark and $j$th quark.

\subsection{Doubly heavy systems in GEM} 

\begin{table*}[!htbp]
\centering
\caption{Three-body and four-body angular-momentum space of $QQq$ and $QQ\bar{q}\bar{q}$. The units of $\nu_1$, $\lambda_1$, and $\omega_1$ are $\text{GeV}^{-2}$, and the remaining quantities are dimensionless. \label{tab:Angular-momentum space}}

\begin{minipage}{0.95\textwidth}
\subcaption{ $QQq$ systems}
\centering
\begin{tabular*}{\textwidth}{@{\extracolsep{\fill}}cccccccccccc}
    \toprule[1pt]
     \toprule[1.00pt]
    $J^P$ & $l$ & $L$ & $L_t$ & $s_{23}$ & $S$ & $n_{\max}$ & $\nu_1$ & $q$
    & $N_{\max}$ & $\lambda_1$ & $Q$ \\
    \midrule
    $\frac{1}{2}^+$ & 0 & 0 & 0 & 1 & $\tfrac12$ & 20 & 20 & 0.6 & 20 & 30 & 0.6 \\
    $\frac{3}{2}^+$ & 0 & 0 & 0 & 1 & $\tfrac32$ & 20 & 20 & 0.6 & 20 & 30 & 0.6 \\
    \bottomrule[1pt]
    \bottomrule[1pt]
\end{tabular*}
\end{minipage}

\vspace{1.5em}

\begin{minipage}{0.95\textwidth}
\subcaption{ $QQ\bar q\bar q$ systems}
\centering
\begin{tabular*}{\textwidth}{@{\extracolsep{\fill}}cccccccccccccccccc}
    \toprule[1pt]
    \toprule[1pt]
    $J^P$ & $l$ & $L$ & $I$ & $\lambda$ & $L_t$ & $s_{12}$ & $s_{34}$ & $S$
    & $n_{\max}$ & $\nu_1$ & $q$ & $N_{\max}$ & $\lambda_1$ & $Q$
    & $\nu_{\max}$ & $\omega_1$ & $a$ \\
    \midrule
    $0(1^+)$ & 0 & 0 & 0 & 0 & 0 & 1 & 0 & 1 & 6 & 0.3 & 0.1 & 6 & 0.2 & 0.1 & 6 & 0.1 & 0.1 \\
    $1(0^+)$ & 0 & 0 & 0 & 0 & 0 & 1 & 1 & 0 & 6 & 0.3 & 0.1 & 6 & 0.2 & 0.1 & 6 & 0.1 & 0.1 \\
    $1(1^+)$ & 0 & 0 & 0 & 0 & 0 & 1 & 1 & 1 & 6 & 0.3 & 0.1 & 6 & 0.2 & 0.1 & 6 & 0.1 & 0.1 \\
    $1(2^+)$ & 0 & 0 & 0 & 0 & 0 & 1 & 1 & 2 & 6 & 0.3 & 0.1 & 6 & 0.2 & 0.1 & 6 & 0.1 & 0.1 \\
    \bottomrule[1pt]
    \bottomrule[1pt]
\end{tabular*}
\end{minipage}

\end{table*}

Following the approach of our previous work \cite{Luo:2023sne,Man:2024mvl,Zhou:2025fpp,Luo:2025psq,Zhang:2025vtc,Luo:2025cqs,An:2025rjv,Luo:2023sra,Peng:2024pyl,Luo:2025cqs,Zhou:2025fpp}, we use a well-chosen set of Gaussian basis functions forming an approximate complete set in a finite coordinate space to expand the wave function and transform the stationary Schr\"{o}dinger equation into a generalized eigenvalue problem. 

For the three-body $QQq$ systems, with two heavy quarks $Q$ positioned at $m_2$ and $m_3$ in Fig.~1 of Ref.~\cite{Hiyama:2003cu}, we use the symmetric basis
\begin{equation}\label{eq:Phi}
    \Phi_\alpha = \left[[\phi_{nl}^{\mathrm{G}}(\boldsymbol{r}_1)\phi_{NL}^{\mathrm{G}}(\boldsymbol{R}_1)]_{L_t}[s_1[s_2s_3]_{s_{23}}]_S\right]_{JM},
\end{equation}
where $\alpha = \{l, L, L_t, s_{23}, S, n, N\}$. For the four-body $QQ\bar{q}\bar{q}$ systems, with two heavy quarks $Q$ positioned at $m_1$ and $m_2$ and two light quarks $\bar{q}$ positioned at $m_3$ and $m_4$ in Fig.~18 of Ref.~\cite{Hiyama:2003cu}, we use the basis
\begin{equation}\label{eq:Phi2}
    \Phi_\alpha = \left[[[\phi_{nl}^{\mathrm{G}}(\boldsymbol{r}_c)\phi_{NL}^{\mathrm{G}}(\boldsymbol{R}_c)]_I\phi_{\nu\lambda}^{\mathrm{G}}(\boldsymbol{\rho}_c)]_{L_t}[[s_1s_2]_{s_{12}}[s_3s_4]_{s_{34}}]_S\right]_{JM},
\end{equation}
where $\alpha = \{c, l, L, I, \lambda, L_t, s_{12}, s_{34}, S, n, N, \nu\}$. The $\alpha$ used for each $J^P$ quantum number is shown in Table~\ref{tab:Angular-momentum space}.

To estimate the discussed hadron masses, several input parameters are required.
The coupling constant $\alpha_s(\mathbf{q})$, including the one-loop QCD correction, is given by \cite{Richardson:1978bt,Tang:1995iy}.
\begin{equation}\label{eq:alphas}
\alpha_s(\mathbf{q})=\frac{12\pi}{33-2N_f}\frac{1}{\text{log}(e+\frac{\mathbf{q}^2}{\Lambda^2})},
\end{equation}
where $N_f=5$, $e=2.71828$, and $\Lambda=0.1$ GeV \cite{,Wang:2024hzd}.
The value of $\alpha_s(m_i)$ is then obtained from above expression using the constitute quark mass $m_i$.
For string tension, we use the value $b=0.15$ $\text{GeV}^2$ to calculate the hadron masses, which is consistent with the result from Lattice QCD \cite{Kawanai:2011jt}.

By reproducing the masses of these well-established $S$-wave mesons  using the GEM, as show in Table~\ref{tab:comparisons},
we determine the remaining model parameters, which are summarized in Table \ref{parameter}. 
In Ref.~\cite{Li:2023wug}, the coefficients  $C_{ij}$ of the color-magnetic interaction were determined from the measured masses of $S$-wave hadrons. We therefore adopt the same set of $C_{ij}$ value in the present work.
A comprehensive description of the extracted $C_{ij}$ parameters can be found in Ref.~\cite{Li:2023wug}.
We employ the variational parameters to be $0.5$ GeV, $0.3$ GeV,  and $0.4$ GeV in BOA-STFs for the $QQq$ ($q=u,d,s$), $QQ\bar{n}\bar{n}$ ($n=u,d$), and $QQ\bar{s}\bar{s}$, respectively. 

With the above parameters, the estimated eigenvalues and  masses for doubly heavy baryons and tetraquarks are listed in Tables \ref{baryon}, \ref{ccnn}, and \ref{ccss}.

\begin{table}[!htbp]
    \caption{The comparisons of experimental and theoretical masses. The $M^{\rm The.}$, $M^{\rm Exp.}$, and $M^{\rm Err.}$ are theoretical calculations, experimental results, and uncertainties of the masses, respectively. We also present $\chi^2/n$, where $n$ is the number of the states.}
    \label{tab:comparisons}
    \renewcommand\arraystretch{1.5}
    \begin{tabular*}{86mm}{@{\extracolsep{\fill}}cccc}
        \toprule[1.00pt]
        \toprule[1.00pt]
        States                      &$M^{\rm The.}$ (MeV)&$M^{\rm Exp.}$ (MeV)~\cite{ParticleDataGroup:2024cfk} &$M^{\rm Err.}$ (MeV)~\cite{ParticleDataGroup:2024cfk}\\
        \midrule[0.75pt]
        $D^\pm$                 &1864.4&1869.5&0.4\\
        $D^{*\pm}$             &2010.2&2010.26&0.05\\
        $D_s^\pm$                &1968.5&1968.35&0.07\\
        $D_s^{*\pm}$                &2101.2&2106.6&3.4\\
        $B^\pm$                 &5282.5&5279.41&0.07\\
        $B^*$             &5300.5&5324.75&0.20\\
        $B_s^0$                &5366.8&5366.91&0.11\\
        $B_s^*$                &5385.8&5415.4&1.4\\
        \multicolumn{4}{c}{$\chi^2/n=2155.7$}\\
        \bottomrule[1.00pt]
        \bottomrule[1.00pt]
    \end{tabular*}
\end{table}

\begin{table}[!htbp]
\centering
\caption{Comparison of model parameters used in the GEM and BOA. 
The quark masses and confinement parameters are identical in both approaches. 
The hyperfine part differs: the potential model uses a single parameter $\sigma$, 
while the BOA employs twelve coefficients $C_{ij}$.}
\renewcommand{\arraystretch}{1.2}\label{parameter}
\begin{tabular}{cccc}
 \toprule[1.00pt]
\toprule[1.00pt]
& & GEM & BOA \\ 
\hline
Quark masses&\quad $m_{n}$(GeV) & 0.3971 & same \\
&\quad $m_{s}$(MeV) & 576.1 & same \\
&\quad $m_{c}$(MeV) & 1598.8 & same \\
&\quad $m_{b}$(MeV) & 4893.1 & same \\
&\quad $m_{t}$(MeV) & 172560 & same \\
\hline
Confinement&\quad $\alpha_s$ & Eq.~(\ref{eq:alphas}) & same \\
&\quad $b$ (GeV$^2$) & 0.15 & same \\
&\quad $C$ (MeV) & -606.9 & same \\
\hline
Hyperfine&\quad $\sigma$ (MeV) & 3772.2 & --- \\
&\quad $C_{nn}$(MeV) & --- & 18.3 \\
&\quad $C_{cc}$(MeV)  & --- & 3.5 \\
&\quad $C_{cn}$(MeV)  & --- & 4.0 \\
&\quad $C_{cs}$(MeV)  & --- & 4.3 \\
&\quad $C_{bb}$(MeV)  & --- & 1.9 \\
&\quad $C_{bn}$(MeV)  & --- & 1.3 \\
&\quad $C_{bs}$(MeV)  & --- & 1.3 \\
&\quad $C_{c\bar{n}}$(MeV)  & --- & 6.6 \\
&\quad $C_{b\bar{n}}$(MeV)  & --- & 2.1 \\
&\quad $C_{ss}$ (MeV) & --- & 6.5 \\
&\quad $C_{c\bar{s}}$(MeV)  & --- & 6.7 \\
&\quad $C_{b\bar{s}}$(MeV)  & --- & 2.3 \\
\bottomrule[1.00pt]
\bottomrule[1.00pt]
\end{tabular}
\end{table}

\subsection{Doubly heavy baryon}

\begin{figure*}
\centering
\begin{tabular*}{0.9\textwidth}{@{\extracolsep{\fill}}cc}
\scalebox{0.75}{
    \begin{tikzpicture}
		\begin{axis}[  
   tick label style={font=\normalsize},
    label style={font=\normalsize},
    legend style={font=\normalsize},
			xlabel=$m_Q$ (GeV),
			xmin=1.4, xmax=10, 
			ylabel=$E$ (GeV),
       legend style={
        at={(0.7,0.97)},   
        anchor=north east,  
        draw=none,          
        fill=none,          
        nodes={scale=0.8, anchor=west}, 
        cells={anchor=west},           
        column sep=5pt,                
        font=\normalsize                   
        },
   			legend entries={GEM, BOA-STFs, BOA-GTFs},
			legend pos=north east,
			mark size=1pt,
			set layers,
			]
      \draw[black!50, thick, dashed]
  (axis cs:1.6,-1.5) -- (axis cs:1.6,1);
  \node[anchor=west,font=\normalsize] at (axis cs:1.6,-0.25) {$m_c$};
  \draw[black!50, thick, dashed]
  (axis cs:4.9,-1.5) -- (axis cs:4.9,1);
 \node[anchor=west,font=\normalsize] at (axis cs:4.9,-0.25) {$m_b$};
			\addplot+coordinates{

(1.50000000,-0.00021897)
(2.00000000,-0.03691253)
(2.50000000,-0.06305877)
(3.00000000,-0.08307086)
(3.50000000,-0.09915129)
(4.00000000,-0.11244851)
(4.50000000,-0.12378052)
(5.00000000,-0.13360114)
(5.50000000,-0.14225805)
(6.00000000,-0.14998061)
(6.50000000,-0.15692954)
(7.00000000,-0.16326475)
(7.50000000,-0.16907128)
(8.00000000,-0.17441505)
(8.50000000,-0.17939191)
(9.00000000,-0.18403869)
(9.50000000,-0.18838674)
(10.00000000,-0.19246210)
			};
            \addplot+coordinates{
(1.5, -0.00409129)
(2., -0.0336974)
(2.5, -0.0554299)
(3., -0.0723627)
(3.5, -0.0861118)
(4., -0.097618)
(4.5, -0.107471)
(5., -0.116062)
(5.5, -0.123663)
(6., -0.130467)
(6.5, -0.13662)
(7., -0.142232)
(7.5, -0.147386)
(8., -0.152151)
(8.5, -0.15658)
(9., -0.160718)
(9.5, -0.164599)
(10., -0.168256)
             };
              \addplot+
              [ color=green!70!black,
    mark=triangle*,
    mark size=1.5pt] 
coordinates{(1.5, -0.0274784)
(2., -0.0725344)
(2.5, -0.104827)
(3., -0.129643)
(3.5, -0.149612)
(4., -0.166216)
(4.5, -0.180364)
(5., -0.192651)
(5.5, -0.203482)
(6., -0.21315)
(6.5, -0.221868)
(7., -0.229797)
(7.5, -0.237063)
(8., -0.243763)
(8.5, -0.249977)
(9., -0.255769)
(9.5, -0.26119)
(10., -0.266285)
              };
		\end{axis}
	\end{tikzpicture}}

& 
    \scalebox{0.75}{
    \begin{tikzpicture}
		\begin{axis}[
    tick label style={font=\normalsize},
    label style={font=\normalsize},
    legend style={font=\normalsize},
			xlabel=$m_Q$ (GeV),
			xmin=1.4, xmax=10, 
			ylabel=$E$ (GeV),
       legend style={
        at={(0.7,0.97)},   
        anchor=north east,  
        draw=none,          
        fill=none,          
        nodes={scale=0.9, anchor=west}, 
        cells={anchor=west},           
        column sep=5pt,                
        font=\small                    
        },
			legend entries={GEM, BOA-STFs,BOA-GTFs},
			legend pos=north east,
			mark size=1pt,
			set layers]
      \draw[black!50, thick, dashed]
  (axis cs:1.6,-1.5) -- (axis cs:1.6,1);
  \node[anchor=west,font=\normalsize] at (axis cs:1.6,-0.03) {$m_c$};
  \draw[black!50, thick, dashed]
  (axis cs:4.9,-1.5) -- (axis cs:4.9,1);
 \node[anchor=west,font=\normalsize] at (axis cs:4.9,-0.03) {$m_b$};
			\addplot+coordinates{
(1.50000000,0.21087376)
(2.00000000,0.17603514)
(2.50000000,0.15059791)
(3.00000000,0.13113739)
(3.50000000,0.11577618)
(4.00000000,0.10322431)
(4.50000000,0.09268639)
(5.00000000,0.08366142)
(5.50000000,0.07579774)
(6.00000000,0.06884561)
(6.50000000,0.06262279)
(7.00000000,0.05699305)
(7.50000000,0.05185261)
(8.00000000,0.04712104)
(8.50000000,0.04273507)
(9.00000000,0.03864426)
(9.50000000,0.03480779)
(10.00000000,0.03119208)
			};
            \addplot+coordinates{
(1.5, 0.240374)
(2., 0.210981)
(2.5, 0.189172)
(3., 0.172031)
(3.5, 0.158012)
(4., 0.146208)
(4.5, 0.136048)
(5., 0.127148)
(5.5, 0.119243)
(6., 0.112141)
(6.5, 0.105698)
(7., 0.0998068)
(7.5, 0.0943812)
(8., 0.0893544)
(8.5, 0.0846722)
(9., 0.0802905)
(9.5, 0.0761729)
(10., 0.0722889)
             };
\addplot+[ color=green!70!black,
    mark=triangle*,
    mark size=1.5pt]  coordinates{
    (1.5, 0.19989)
(2., 0.153315)
(2.5, 0.119578)
(3., 0.0934273)
(3.5, 0.0722356)
(4., 0.0545131)
(4.5, 0.0393403)
(5., 0.0261115)
(5.5, 0.0144088)
(6., 0.00393197)
(6.5, -0.00554077)
(7., -0.0141777)
(7.5, -0.0221093)
(8., -0.0294384)
(8.5, -0.0362474)
(9., -0.0426035)
(9.5, -0.0485619)
(10., -0.0541686)
};
		\end{axis}
	\end{tikzpicture}}\\
\hspace{1.75em}(a)&\hspace{1.75em}(b)
\end{tabular*}
    \caption{The dependence of the eigenvalues $E$ on the heavy quark mass $m_Q$ with GEM (blue line), BOA-STFs (red line), and BOA-GTFs (green line) for $QQn$ and $QQnn$.} \label{ccn}
\end{figure*}
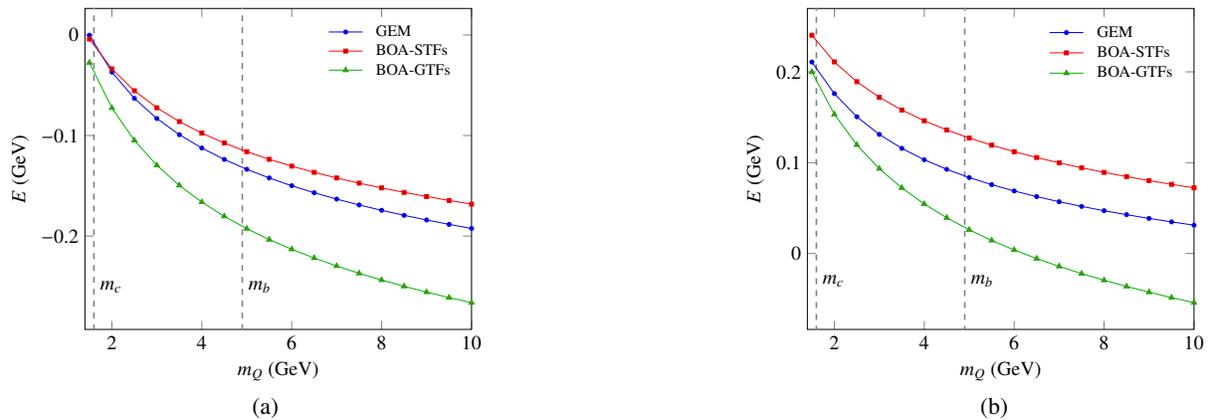

The GEM provides highly accurate solutions to the multi-body Schr\"{o}dinger equation and is therefore well suited as a benchmark for investigating hadronic properties.
Here, to enable a direct comparison the BOA-STFs, BOA-GTFs and GEM eigenvalues, the hyperfine interactions are temporarily neglected and only the kinetic energy, Coulomb and confinement terms are considered. For the $QQn$, the dependence of the eigenvalue  $E$ on the heavy-quark mass $m_Q$ is illustrated in Fig.~\ref{ccn} (a).

As shown in the Fig. \ref{ccn} (a), 
in hadronic systems, the confinement term and the attractive interaction between two heavy quarks lead to similar binding energies when the heavy-quark mass is small.
Furthermore, the eigenvalues  exhibit a decreasing trend with increasing quark mass in three methods, reflecting their dependence on the heavy-quark mass.  

A qualitative difference emerges when comparing the mass dependence of the binding energies in molecular systems and in multiquark states containing two heavy quarks. In molecular systems , the BOA binding energy is entirely determined by the electronic dynamics and remains independent of the nuclear mass. Consequently, in the heavy-mass limit the kinetic energy of the nuclei becomes negligible.
However, for doubly heavy baryons and tetraquark systems, the BOA binding energy depends explicitly on the heavy-quark mass.

With increasing heavy-quark mass, 
the light-quark dynamics in the BOA framework are treated variationally, and the simplified STFs used as trial wave functions tend to enhance the binding energy. This behavior mainly arises from their limited ability to capture the long-range confining potential.

For comparison, GTFs are also employed as trial wave functions within the BOA framework. For a heavy-quark mass equal to the charm-quark mass, the BOA-GTFs yield results smaller than those obtained from GEM. This underestimation in the BOA-GTF calculations mainly originates from the neglect of non-adiabatic corrections inherent in the BOA.

\begin{table}[!h]
\caption{ Eigenvalues and mass spectra for the $QQn$ $(n=u,d)$ and $QQs$ states. All eigenvalues and masses are given in units of MeV. }\centering
\renewcommand\arraystretch{1.35}\label{baryon}
\begin{tabular*}{1.0\columnwidth}{@{\extracolsep{\fill}}c cc cc cc}
\toprule[1.00pt]
\toprule[1.00pt]
\multirow{2}{*}{States} & \multicolumn{2}{c}{BOA-STFs} &  \multicolumn{2}{c}{BOA-GTFs}&\multicolumn{2}{c}{GEM} \\
\cmidrule(lr){2-3} \cmidrule(lr){4-5} \cmidrule(lr){6-7}
& $E_{\text{BOA}}$ & Mass & $E_{\text{BOA}}$ & Mass & $E$ & Mass \\ \midrule
$\Xi_{cc}$        & -10.9   & 3550.5  &-37.8&3525.5 & -7.9 & 3566.4\\
$\Xi_{cc}^{*}$    & -10.9   & 3614.5  &-37.8& 3587.4& -7.9 & 3600.3\\
$\Xi_{bb}$        & -114.4 & 10060.1  &-190.2&9984.4 & -131.4 & 10046.5\\
$\Xi_{bb}^*$      & -114.4 & 10080.9  &-190.2&10005.1 & -131.4 & 10055.9\\
$\Omega_{cc}$     & -94.4  & 3642.8   &-134.0&3603.2 & -95.0 & 3660.1\\
$\Omega_{cc}^{*}$ & -94.4  & 3711.6   &-134.0&3672.0 & -95.0 & 3691.8\\
$\Omega_{bb}$     & -202.3 & 10151.2  &-292.6&10060.9 & -223.7 & 10133.3\\
$\Omega_{bb}^{*}$ & -202.3 & 10172.0  &-292.6&10081.7 & -223.7 & 10142.6\\
\bottomrule[1.00pt]
\bottomrule[1.00pt]
\end{tabular*}
\end{table}

From Table. \ref{baryon}, after including the hyperfine interactions, the masses of the $\Xi_{cc}$ obtained using the  BOA-STFs and BOA-GTFs are found to be 3550.5  MeV and 3525.5 MeV, respectively. 
The corresponding eigenvalues, $E=-10.9$ MeV and $E=-37.8$ MeV, which are close to the results obtained from the GEM. For the $\Xi_{bb}$ system, the eigenvalue predicted by the BOA-STFs is $-114.4$~MeV, which is larger than the corresponding GEM result. In contrast, the BOA-GTFs approach yields a smaller eigenvalue of $-190.2$~MeV compared with the GEM calculation.
The predicted masses and eigenvalues  of the remaining doubly heavy baryons are also listed in  Table. \ref{baryon}.

\subsection{Doubly heavy tetraquark: $QQ$ in color $\bar{3}$}
For the $QQnn$ tetraquark states,  Fig.~\ref{ccn} (b) shows the variation of the  $E$ with respect to the heavy quark mass, as obtained from the BOA-STFs, BOA-GTFs, and  GEM.

As shown in Fig.~\ref{ccn} (b), the variation trend of the eigenvalues for the doubly heavy tetraquark states is similar to that for doubly heavy baryons.
For small heavy-quark masses, the binding energies predicted by the BOA are close to those obtained from the GEM. However, as the heavy-quark mass increases, the BOA-STFs predictions for the $QQ\bar{n}\bar{n}$ states become larger than those obtained from the GEM, whereas the BOA-GTFs results are smaller than the GEM ones.

\begin{table}[!htbp]
\caption{ Eigenvalues and mass spectra for the $cc\bar{n}\bar{n}$ and $bb\bar{n}\bar{n}$ states. All eigenvalues and masses are given in units of MeV. }\centering
\renewcommand\arraystretch{1.35}\label{ccnn}
\begin{tabular*}{1.0\columnwidth}{@{\extracolsep{\fill}}c cc cc cc}
\toprule[1.00pt]
\toprule[1.00pt]
\multicolumn{7}{c}{$cc\bar{n}\bar{n}$ system}\\\hline
\multirow{2}{*}{$I(J^P)$} & \multicolumn{2}{c}{BOA-STFs}& \multicolumn{2}{c}{BOA-GTFs} & \multicolumn{2}{c}{GEM} \\
\cmidrule(lr){2-3} \cmidrule(lr){4-5} \cmidrule(lr){6-7}
& $E_{\text{BOA}}$ & Mass & $E_{\text{BOA}}$ & Mass & $E$ & Mass  \\ \midrule
$0(1^+)$ & 233.7  & 4088.4  &188.5&4043.2 & 195.8 & 4051.1\\
$1(0^+)$ & 233.7  & 4213.2  &188.5&4168.0 & 195.8  & 4184.7\\
$1(1^+)$ & 233.7  & 4248.4  &188.5&4203.2 & 195.8  & 4193.3\\
$1(2^+)$ & 233.7 &  4318.8  &188.5&4273.6 & 195.8  & 4208.8\\ \hline\hline
\multicolumn{7}{c}{$bb\bar{n}\bar{n}$ system}\\ \hline
$0(1^+)$ & 128.9   & 10568.0 &27.7&10466.7 & 84.4 & 10615.4\\
$1(0^+)$ & 128.9   & 10740.8 &27.7&10639.5 &84.4 & 10668.7\\
$1(1^+)$ & 128.9   & 10752.0  &27.7&10650.7&84.4 & 10670.9\\
$1(2^+)$ & 128.9   & 10774.4  &27.7&10673.1&84.4 & 10675.1\\
\bottomrule[1.00pt]
\bottomrule[1.00pt]
\end{tabular*}
\end{table}

\begin{table}[!htbp]
\caption{ Eigenvalues and mass spectra for the $cc\bar{s}\bar{s}$ and $bb\bar{s}\bar{s}$ states. The eigenvalues and masses are in units of MeV.}\centering\label{ccss}
\begin{tabular*}{1.0\columnwidth}{@{\extracolsep{\fill}}c cc cc cc}
\toprule[1.00pt]
\toprule[1.00pt]
\multicolumn{7}{c}{$cc\bar{s}\bar{s}$ system}\\\hline
\multirow{2}{*}{$J^P$} & \multicolumn{2}{c}{BOA-STFs} & \multicolumn{2}{c}{BOA-GTFs}&\multicolumn{2}{c}{GEM} \\
\cmidrule(lr){2-3} \cmidrule(lr){4-5} \cmidrule(lr){6-7}
& $E_{\text{BOA}}$ & Mass & $E_{\text{BOA}}$ & Mass & $E$ & Mass\\ \midrule
$0^+$ & 61.6   & 4366.6 &18.3&4323.3  & 38.6 & 4384.1\\
$1^+$ & 61.6  & 4402.4  &18.3&4359.1 & 38.6 & 4392.3\\
$2^+$ & 61.6    & 4473.8  &18.3&4430.5 & 38.6 & 4407.0\\\hline\hline
\multicolumn{7}{c}{$bb\bar{s}\bar{s}$ system}\\ \hline
$0^+$ & -38.4  & 10897.9 &-136.8&10799.5 & -76.2 & 10864.3\\
$1^+$ & -38.4  & 10910.1 &-136.8&10811.8 & -76.2 & 10866.5\\
$2^+$ & -38.4   & 10934.7 &-136.8& 10836.3& -76.2 & 10870.8\\
\bottomrule[1.00pt]
\bottomrule[1.00pt]
\end{tabular*}
\end{table}

From Table~\ref{ccnn},  the  mass of the $I(J^P)=0(1^+)$ $cc\bar{n}\bar{n}$ state predicted within the BOA-STFs  and BOA-GTFs  are 4088.4~\text{MeV} and 4043.2~\text{MeV}, respectively. Both values are close to the corresponding result obtained using the  GEM (4051.1~\text{MeV})
The  masses of the isovector ($I=1$) $cc\bar{n}\bar{n}$ states are found to lie around  4.2 GeV. The mass splitting between the isoscalar and isovector $cc\bar{n}\bar{n}$ states is primarily determined by the hyperfine interaction of the Hamiltonian.

From Table~\ref{ccnn}, the eigenvalues of the $I(J^{P})=0(1^{+})$ $bb\bar{n}\bar{n}$ state predicted within the BOA-STFs and BOA-GTFs frameworks are 128.9 MeV and 27.7 MeV, respectively. The BOA-STFs result indicates a weaker binding than the GEM, whereas the BOA-GTFs approach leads to a stronger binding.
Other mass estimates for the doubly heavy tetraquark states, obtained using different approaches, are listed in Tables~\ref{ccnn} and~\ref{ccss}.

\section{Summary}\label{secV}

In this work, we investigate the masses and binding properties of hydrogen-like systems using both the BOA and the GEM frameworks.
To examine the validity of the BOA
within potential models, we further study doubly heavy baryons and
tetraquarks within the same frameworks.
Since the  GEM provides highly reliable solutions to the full multi-body  Schr\"{o}dinger equation without assuming a separation of scales,  it is employed as a benchmark for evaluating the applicability of the BOA within potential models. Additionally, we also examine how different choices of trial wave function within the BOA lead to different results.

By analyzing hydrogen-like systems, we demonstrate that the reliability of the BOA is governed by the mass ratio between the heavy and light constituents. When a clear separation of scales is present, the BOA yields results close to the solutions obtained with the GEM. 

However, in hadronic systems with relatively small heavy-quark masses, the three approaches yield comparable results. Significant deviations arise in doubly heavy hadrons, where both the confinement interaction and the attractive interaction between the two heavy quarks play important roles.
As the heavy-quark mass increases, the BOA-STFs predictions tend to exceed those obtained from the GEM, whereas the BOA-GTFs calculations systematically yield lower values. Our analysis indicates that the underestimation in the BOA-GTFs results primarily originates from the neglect of non-adiabatic contributions. In contrast, the use of STFs as trial wave functions tends to overestimate the binding energy, primarily because STFs are not well suited to describe the long-range behavior induced by the confining interaction.

Although the BOA provides an intuitive physical picture for doubly heavy hadrons, its quantitative accuracy is sensitive to the choice of basis functions and the heavy-quark mass scale. 
For doubly heavy systems with charm quarks, BOA can provide reasonable estimates at the qualitative level. However, for bottom and heavier systems, quantitative predictions become less reliable, and a fully dynamical treatment such as GEM is indispensable.

Our results suggest that the conclusions are sensitive to the form of the interaction potential. Since the potential adopted in the present work differs from that used in Refs. ~\cite{Maiani:2019cwl,Liu:2025jyn}, the resulting predictions are not identical to those obtained in those studies. Furthermore,
the present conclusions are specific to the potential model framework.
They should not be interpreted as a general assessment of the validity of the BOA in QCD-based studies ~\cite{Juge:1999ie,Braaten:2014qka,Bicudo:2015vta,Bicudo:2016ooe,Bruschini:2023zkb,Berwein:2024ztx,Braaten:2024stn,Braaten:2024tbm}.

\section*{Acknowledgements}
Zi-Long Man is grateful to Liang-Zhen Wen for helpful discussions.
This work is also supported by the National Natural Science Foundation of China under Grant Nos. 12335001 and 12247101, the ‘111 Center’ under Grant No. B20063, the Natural Science Foundation of Gansu Province (No. 22JR5RA389, No. 25JRRA799), the fundamental Research Funds for the Central Universities, the project for top-notch innovative talents of Gansu province, and Lanzhou City High-Level Talent Funding.

\section*{Appendix:  variational wave functions}
Within the BOA framework, two types of variational wave functions are commonly employed, namely Slater-type functions (STFs) and Gaussian-type functions (GTFs).

We first employ STFs to estimate the masses of doubly  heavy baryons and doubly  heavy tetraquark states.
The wave function of the light quark in the doubly heavy baryon system can be expressed as 
\begin{equation}
	\Phi=\mathcal{C}[R(r_A)+R(r_B)].
\end{equation}
where the radial variation wave function $R(r)$ is Slater-type functions, which has the form 
\begin{equation}
	R(r)=\frac{A^{3/2}}{\sqrt{\pi}}e^{-Ar}.
\end{equation}
Here, $A$ is  the variational parameter.
$\mathcal{C}=\frac{1}{\sqrt{2+2I_1}}$ is normalization factor, and $I_1=\int R(r_A)R(r_B)d\tau$.

For the doubly heavy baryon,the eigenvalue associated with the light degree of freedom can be obtained as
\begin{equation}
\begin{split}
\varepsilon(r_{AB})&=\frac{1}{1+I_1}
 \bigg [
 \frac{A^2}{2m_{q}}(1-I_1)-(\frac{4\alpha_s}{3}-\frac{A}{m_{q}})I_2 \\
 &\quad-\frac{2\alpha_s}{3}(A+I_3) +\frac12 b(\frac{3}{2A}+I_4+I_5)  \bigg ]+C,
\end{split}
\end{equation}
where
\begin{widetext}
\begin{gather}
I_2=\int R(r_A)R(r_B)\frac{1}{r_A}d\tau=\int R(r_A)R(r_B)\frac{1}{r_B}d \tau, \\
I_3=\int R(r_B)^2\frac{1}{r_A}d\tau=\int R(r_A)^2\frac{1}{r_B}d\tau,\\
I_4=\int R(r_B)^2r_Ad\tau=\int R(r_A)^2r_Bd\tau,\\
I_5=\int R(r_A)R(r_B)r_Ad\tau=\int R(r_A)R(r_B)r_B d \tau.
\end{gather}

For the doubly heavy tetraquarks, the STFs associated with the light quarks is chosen as 
\begin{equation}
\Phi=\frac{1}{2+2I_1^2}\left(R(r_{A1})R(r_{B2})+R(r_{A2})R(r_{B1}) \right).
\end{equation}
The eigenvalue $\varepsilon(r_{AB})$ of $H_{light}$ is determined from
\begin{equation}
H_{light}\Phi=\varepsilon(r_{AB})\Phi,
\end{equation}
where
\begin{equation}
\begin{split}
\varepsilon(r_{AB})&=
\frac{1}{1+I_1^2}\Bigg \{2\left[\frac{A}{2m_{q}}-\frac{ A \alpha_s}{3}
\left(-\frac{A^2}{2m_{q}}I_1+\frac{A}{m_{q}}I_2-\frac{\alpha_s}{3}I_2
\right)I_1 \right] 
\quad-\frac{2\alpha_s}{3}(I_1I_2+I_3)
-\frac{2\alpha_l}{3}(I_6+I_7)
\quad+\frac{1}{2}bI_1I_5\left(\frac{3}{2A}+I_4\right)\Bigg \}+C.
\end{split}
\end{equation}
\end{widetext}
 Here, the two additional integrals $I_6$ and $I_7$ are defined as
\begin{gather}
I_6=\langle R(r_{A1})R(r_{B2})\frac{1}{r_l} R(r_{A1})R(r_{B2})\rangle, \\
I_7= \langle R(r_{A1})R(r_{B2})\frac{1}{r_l} R(r_{A2})R(r_{B1})  \rangle.
\end{gather}

Similar to STFs, GTFs can also be employed as variational wave functions. For doubly  heavy baryons, the trial wave function is constructed as 
\begin{equation}
\Phi=\sum_{n=1}^{n_{max}}C_{n}\left (\phi_n^{G}(\mathbf{r_A})+\phi_n^{G}(\mathbf{r_B})\right),
\end{equation}
where 
\begin{equation}
\phi_n^G(\mathbf{r})=\frac{1}{4\pi}N_n \text{e}^{-\nu_n r^2},~ N_n=\left(\frac{4(2\nu_n)^{3/2}} {\sqrt{\pi}}\right)^{1/2}.
\end{equation}

The $\nu_n$ is the Gaussian size parameters, which is defined as
\begin{equation}
\nu_n=\frac{1}{r_n^2},~r_n=r_1 a^{n-1}, ~(n=1,2,\cdots,n_{\max}).
\end{equation}

For doubly  heavy tetraquark states, the corresponding wave function is given by
\begin{equation}
\Phi=\sum_{n=1}^{n_{max}}C_{n}\left (\phi_n^{G}(\mathbf{r_{A1}})\phi_n^{G}(\mathbf{r_{B2}})+\phi_n^{G}(\mathbf{r_{A2}})\phi_n^{G}(\mathbf{r_{B1}})\right).
\end{equation}
\bibliography{inport}

\begin{thebibliography}{62}%
\makeatletter
\providecommand \@ifxundefined [1]{%
 \@ifx{#1\undefined}
}%
\providecommand \@ifnum [1]{%
 \ifnum #1\expandafter \@firstoftwo
 \else \expandafter \@secondoftwo
 \fi
}%
\providecommand \@ifx [1]{%
 \ifx #1\expandafter \@firstoftwo
 \else \expandafter \@secondoftwo
 \fi
}%
\providecommand \natexlab [1]{#1}%
\providecommand \enquote  [1]{``#1''}%
\providecommand \bibnamefont  [1]{#1}%
\providecommand \bibfnamefont [1]{#1}%
\providecommand \citenamefont [1]{#1}%
\providecommand \href@noop [0]{\@secondoftwo}%
\providecommand \href [0]{\begingroup \@sanitize@url \@href}%
\providecommand \@href[1]{\@@startlink{#1}\@@href}%
\providecommand \@@href[1]{\endgroup#1\@@endlink}%
\providecommand \@sanitize@url [0]{\catcode `\\12\catcode `\$12\catcode
  `\&12\catcode `\#12\catcode `\^12\catcode `\_12\catcode `\%12\relax}%
\providecommand \@@startlink[1]{}%
\providecommand \@@endlink[0]{}%
\providecommand \url  [0]{\begingroup\@sanitize@url \@url }%
\providecommand \@url [1]{\endgroup\@href {#1}{\urlprefix }}%
\providecommand \urlprefix  [0]{URL }%
\providecommand \Eprint [0]{\href }%
\providecommand \doibase [0]{https://doi.org/}%
\providecommand \selectlanguage [0]{\@gobble}%
\providecommand \bibinfo  [0]{\@secondoftwo}%
\providecommand \bibfield  [0]{\@secondoftwo}%
\providecommand \translation [1]{[#1]}%
\providecommand \BibitemOpen [0]{}%
\providecommand \bibitemStop [0]{}%
\providecommand \bibitemNoStop [0]{.\EOS\space}%
\providecommand \EOS [0]{\spacefactor3000\relax}%
\providecommand \BibitemShut  [1]{\csname bibitem#1\endcsname}%
\let\auto@bib@innerbib\@empty
\bibitem [{\citenamefont {Mattson}\ \emph {et~al.}(2002)\citenamefont {Mattson}
  \emph {et~al.}}]{SELEX:2002wqn}%
  \BibitemOpen
  \bibfield  {author} {\bibinfo {author} {\bibfnamefont {M.}~\bibnamefont
  {Mattson}} \emph {et~al.} (\bibinfo {collaboration} {SELEX}),\ }\bibfield
  {title} {\bibinfo {title} {{First Observation of the Doubly Charmed Baryon
  $\Xi^+_{cc}$}},\ }\href {https://doi.org/10.1103/PhysRevLett.89.112001}
  {\bibfield  {journal} {\bibinfo  {journal} {Phys. Rev. Lett.}\ }\textbf
  {\bibinfo {volume} {89}},\ \bibinfo {pages} {112001} (\bibinfo {year}
  {2002})},\ \Eprint {https://arxiv.org/abs/hep-ex/0208014}
  {arXiv:hep-ex/0208014} \BibitemShut {NoStop}%
\bibitem [{\citenamefont {Ratti}(2003)}]{Ratti:2003ez}%
  \BibitemOpen
  \bibfield  {author} {\bibinfo {author} {\bibfnamefont {S.~P.}\ \bibnamefont
  {Ratti}},\ }\bibfield  {title} {\bibinfo {title} {{New Results on c-Baryons
  and a Search for cc-Baryons in FOCUS}},\ }\href
  {https://doi.org/10.1016/S0920-5632(02)01948-5} {\bibfield  {journal}
  {\bibinfo  {journal} {Nucl. Phys. B Proc. Suppl.}\ }\textbf {\bibinfo
  {volume} {115}},\ \bibinfo {pages} {33} (\bibinfo {year} {2003})}\BibitemShut
  {NoStop}%
\bibitem [{\citenamefont {Aubert}\ \emph {et~al.}(2006)\citenamefont {Aubert}
  \emph {et~al.}}]{BaBar:2006bab}%
  \BibitemOpen
  \bibfield  {author} {\bibinfo {author} {\bibfnamefont {B.}~\bibnamefont
  {Aubert}} \emph {et~al.} (\bibinfo {collaboration} {BaBar}),\ }\bibfield
  {title} {\bibinfo {title} {{Search for doubly charmed baryons $\Xi_{cc}^{+}$
  and $\Xi_{cc}^{++}$ in BABAR}},\ }\href
  {https://doi.org/10.1103/PhysRevD.74.011103} {\bibfield  {journal} {\bibinfo
  {journal} {Phys. Rev. D}\ }\textbf {\bibinfo {volume} {74}},\ \bibinfo
  {pages} {011103} (\bibinfo {year} {2006})},\ \Eprint
  {https://arxiv.org/abs/hep-ex/0605075} {arXiv:hep-ex/0605075} \BibitemShut
  {NoStop}%
\bibitem [{\citenamefont {Chistov}\ \emph {et~al.}(2006)\citenamefont {Chistov}
  \emph {et~al.}}]{Belle:2006edu}%
  \BibitemOpen
  \bibfield  {author} {\bibinfo {author} {\bibfnamefont {R.}~\bibnamefont
  {Chistov}} \emph {et~al.} (\bibinfo {collaboration} {Belle}),\ }\bibfield
  {title} {\bibinfo {title} {{Observation of new states decaying into
  $\Lambda_c^+ K^- \pi^+$ and $\Lambda_c^+ K^0_S \pi^-$}},\ }\href
  {https://doi.org/10.1103/PhysRevLett.97.162001} {\bibfield  {journal}
  {\bibinfo  {journal} {Phys. Rev. Lett.}\ }\textbf {\bibinfo {volume} {97}},\
  \bibinfo {pages} {162001} (\bibinfo {year} {2006})},\ \Eprint
  {https://arxiv.org/abs/hep-ex/0606051} {arXiv:hep-ex/0606051} \BibitemShut
  {NoStop}%
\bibitem [{\citenamefont {Aaij}\ \emph {et~al.}(2020)\citenamefont {Aaij} \emph
  {et~al.}}]{LHCb:2019epo}%
  \BibitemOpen
  \bibfield  {author} {\bibinfo {author} {\bibfnamefont {R.}~\bibnamefont
  {Aaij}} \emph {et~al.} (\bibinfo {collaboration} {LHCb}),\ }\bibfield
  {title} {\bibinfo {title} {{Precision measurement of the $\Xi_{cc}^{++}$
  mass}},\ }\href {https://doi.org/10.1007/JHEP02(2020)049} {\bibfield
  {journal} {\bibinfo  {journal} {JHEP}\ }\textbf {\bibinfo {volume} {02}},\
  \bibinfo {pages} {049}},\ \Eprint {https://arxiv.org/abs/1911.08594}
  {arXiv:1911.08594 [hep-ex]} \BibitemShut {NoStop}%
\bibitem [{\citenamefont {Aaij}\ \emph
  {et~al.}(2018{\natexlab{a}})\citenamefont {Aaij} \emph
  {et~al.}}]{LHCb:2018zpl}%
  \BibitemOpen
  \bibfield  {author} {\bibinfo {author} {\bibfnamefont {R.}~\bibnamefont
  {Aaij}} \emph {et~al.} (\bibinfo {collaboration} {LHCb}),\ }\bibfield
  {title} {\bibinfo {title} {{Measurement of the Lifetime of the Doubly Charmed
  Baryon $\Xi_{cc}^{++}$}},\ }\href
  {https://doi.org/10.1103/PhysRevLett.121.052002} {\bibfield  {journal}
  {\bibinfo  {journal} {Phys. Rev. Lett.}\ }\textbf {\bibinfo {volume} {121}},\
  \bibinfo {pages} {052002} (\bibinfo {year} {2018}{\natexlab{a}})},\ \Eprint
  {https://arxiv.org/abs/1806.02744} {arXiv:1806.02744 [hep-ex]} \BibitemShut
  {NoStop}%
\bibitem [{\citenamefont {Aaij}\ \emph
  {et~al.}(2018{\natexlab{b}})\citenamefont {Aaij} \emph
  {et~al.}}]{LHCb:2018pcs}%
  \BibitemOpen
  \bibfield  {author} {\bibinfo {author} {\bibfnamefont {R.}~\bibnamefont
  {Aaij}} \emph {et~al.} (\bibinfo {collaboration} {LHCb}),\ }\bibfield
  {title} {\bibinfo {title} {{First Observation of the Doubly Charmed Baryon
  Decay $\Xi_{cc}^{++}\rightarrow \Xi_{c}^{+}\pi^{+}$}},\ }\href
  {https://doi.org/10.1103/PhysRevLett.121.162002} {\bibfield  {journal}
  {\bibinfo  {journal} {Phys. Rev. Lett.}\ }\textbf {\bibinfo {volume} {121}},\
  \bibinfo {pages} {162002} (\bibinfo {year} {2018}{\natexlab{b}})},\ \Eprint
  {https://arxiv.org/abs/1807.01919} {arXiv:1807.01919 [hep-ex]} \BibitemShut
  {NoStop}%
\bibitem [{\citenamefont {Aaij}\ \emph
  {et~al.}(2022{\natexlab{a}})\citenamefont {Aaij} \emph
  {et~al.}}]{LHCb:2022rpd}%
  \BibitemOpen
  \bibfield  {author} {\bibinfo {author} {\bibfnamefont {R.}~\bibnamefont
  {Aaij}} \emph {et~al.} (\bibinfo {collaboration} {LHCb}),\ }\bibfield
  {title} {\bibinfo {title} {{Observation of the doubly charmed baryon decay $
  {\Xi}_{cc}^{++}\to {\Xi}_c^{\prime +}{\pi}^{+} $}},\ }\href
  {https://doi.org/10.1007/JHEP05(2022)038} {\bibfield  {journal} {\bibinfo
  {journal} {JHEP}\ }\textbf {\bibinfo {volume} {05}},\ \bibinfo {pages}
  {038}},\ \Eprint {https://arxiv.org/abs/2202.05648} {arXiv:2202.05648
  [hep-ex]} \BibitemShut {NoStop}%
\bibitem [{\citenamefont {Aaij}\ \emph {et~al.}(2025)\citenamefont {Aaij} \emph
  {et~al.}}]{LHCb:2025shu}%
  \BibitemOpen
  \bibfield  {author} {\bibinfo {author} {\bibfnamefont {R.}~\bibnamefont
  {Aaij}} \emph {et~al.} (\bibinfo {collaboration} {LHCb}),\ }\bibfield
  {title} {\bibinfo {title} {{Observation of the doubly-charmed-baryon decay $
  {\Xi}_{cc}^{++}\to {\Xi}_c^0{\pi}^{+}{\pi}^{+} $}},\ }\href
  {https://doi.org/10.1007/JHEP10(2025)136} {\bibfield  {journal} {\bibinfo
  {journal} {JHEP}\ }\textbf {\bibinfo {volume} {10}},\ \bibinfo {pages}
  {136}},\ \Eprint {https://arxiv.org/abs/2504.05063} {arXiv:2504.05063
  [hep-ex]} \BibitemShut {NoStop}%
\bibitem [{\citenamefont {Aaij}\ \emph
  {et~al.}(2022{\natexlab{b}})\citenamefont {Aaij} \emph
  {et~al.}}]{LHCb:2021vvq}%
  \BibitemOpen
  \bibfield  {author} {\bibinfo {author} {\bibfnamefont {R.}~\bibnamefont
  {Aaij}} \emph {et~al.} (\bibinfo {collaboration} {LHCb}),\ }\bibfield
  {title} {\bibinfo {title} {{Observation of an exotic narrow doubly charmed
  tetraquark}},\ }\href {https://doi.org/10.1038/s41567-022-01614-y} {\bibfield
   {journal} {\bibinfo  {journal} {Nature Phys.}\ }\textbf {\bibinfo {volume}
  {18}},\ \bibinfo {pages} {751} (\bibinfo {year} {2022}{\natexlab{b}})},\
  \Eprint {https://arxiv.org/abs/2109.01038} {arXiv:2109.01038 [hep-ex]}
  \BibitemShut {NoStop}%
\bibitem [{\citenamefont {Aaij}\ \emph
  {et~al.}(2022{\natexlab{c}})\citenamefont {Aaij} \emph
  {et~al.}}]{LHCb:2021auc}%
  \BibitemOpen
  \bibfield  {author} {\bibinfo {author} {\bibfnamefont {R.}~\bibnamefont
  {Aaij}} \emph {et~al.} (\bibinfo {collaboration} {LHCb}),\ }\bibfield
  {title} {\bibinfo {title} {{Study of the doubly charmed tetraquark
  $T_{cc}^{+}$}},\ }\href {https://doi.org/10.1038/s41467-022-30206-w}
  {\bibfield  {journal} {\bibinfo  {journal} {Nature Commun.}\ }\textbf
  {\bibinfo {volume} {13}},\ \bibinfo {pages} {3351} (\bibinfo {year}
  {2022}{\natexlab{c}})},\ \Eprint {https://arxiv.org/abs/2109.01056}
  {arXiv:2109.01056 [hep-ex]} \BibitemShut {NoStop}%
\bibitem [{\citenamefont {Chen}\ \emph {et~al.}(2016)\citenamefont {Chen},
  \citenamefont {Chen}, \citenamefont {Liu},\ and\ \citenamefont
  {Zhu}}]{Chen:2016qju}%
  \BibitemOpen
  \bibfield  {author} {\bibinfo {author} {\bibfnamefont {H.-X.}\ \bibnamefont
  {Chen}}, \bibinfo {author} {\bibfnamefont {W.}~\bibnamefont {Chen}}, \bibinfo
  {author} {\bibfnamefont {X.}~\bibnamefont {Liu}},\ and\ \bibinfo {author}
  {\bibfnamefont {S.-L.}\ \bibnamefont {Zhu}},\ }\bibfield  {title} {\bibinfo
  {title} {{The hidden-charm pentaquark and tetraquark states}},\ }\href
  {https://doi.org/10.1016/j.physrep.2016.05.004} {\bibfield  {journal}
  {\bibinfo  {journal} {Phys. Rept.}\ }\textbf {\bibinfo {volume} {639}},\
  \bibinfo {pages} {1} (\bibinfo {year} {2016})},\ \Eprint
  {https://arxiv.org/abs/1601.02092} {arXiv:1601.02092 [hep-ph]} \BibitemShut
  {NoStop}%
\bibitem [{\citenamefont {Liu}\ \emph {et~al.}(2019)\citenamefont {Liu},
  \citenamefont {Chen}, \citenamefont {Chen}, \citenamefont {Liu},\ and\
  \citenamefont {Zhu}}]{Liu:2019zoy}%
  \BibitemOpen
  \bibfield  {author} {\bibinfo {author} {\bibfnamefont {Y.-R.}\ \bibnamefont
  {Liu}}, \bibinfo {author} {\bibfnamefont {H.-X.}\ \bibnamefont {Chen}},
  \bibinfo {author} {\bibfnamefont {W.}~\bibnamefont {Chen}}, \bibinfo {author}
  {\bibfnamefont {X.}~\bibnamefont {Liu}},\ and\ \bibinfo {author}
  {\bibfnamefont {S.-L.}\ \bibnamefont {Zhu}},\ }\bibfield  {title} {\bibinfo
  {title} {{Pentaquark and Tetraquark states}},\ }\href
  {https://doi.org/10.1016/j.ppnp.2019.04.003} {\bibfield  {journal} {\bibinfo
  {journal} {Prog. Part. Nucl. Phys.}\ }\textbf {\bibinfo {volume} {107}},\
  \bibinfo {pages} {237} (\bibinfo {year} {2019})},\ \Eprint
  {https://arxiv.org/abs/1903.11976} {arXiv:1903.11976 [hep-ph]} \BibitemShut
  {NoStop}%
\bibitem [{\citenamefont {Chen}\ \emph {et~al.}(2023)\citenamefont {Chen},
  \citenamefont {Chen}, \citenamefont {Liu}, \citenamefont {Liu},\ and\
  \citenamefont {Zhu}}]{Chen:2022asf}%
  \BibitemOpen
  \bibfield  {author} {\bibinfo {author} {\bibfnamefont {H.-X.}\ \bibnamefont
  {Chen}}, \bibinfo {author} {\bibfnamefont {W.}~\bibnamefont {Chen}}, \bibinfo
  {author} {\bibfnamefont {X.}~\bibnamefont {Liu}}, \bibinfo {author}
  {\bibfnamefont {Y.-R.}\ \bibnamefont {Liu}},\ and\ \bibinfo {author}
  {\bibfnamefont {S.-L.}\ \bibnamefont {Zhu}},\ }\bibfield  {title} {\bibinfo
  {title} {{An updated review of the new hadron states}},\ }\href
  {https://doi.org/10.1088/1361-6633/aca3b6} {\bibfield  {journal} {\bibinfo
  {journal} {Rept. Prog. Phys.}\ }\textbf {\bibinfo {volume} {86}},\ \bibinfo
  {pages} {026201} (\bibinfo {year} {2023})},\ \Eprint
  {https://arxiv.org/abs/2204.02649} {arXiv:2204.02649 [hep-ph]} \BibitemShut
  {NoStop}%
\bibitem [{\citenamefont {Liu}\ \emph {et~al.}(2025{\natexlab{a}})\citenamefont
  {Liu}, \citenamefont {Pan}, \citenamefont {Liu}, \citenamefont {Wu},
  \citenamefont {Lu},\ and\ \citenamefont {Geng}}]{Liu:2024uxn}%
  \BibitemOpen
  \bibfield  {author} {\bibinfo {author} {\bibfnamefont {M.-Z.}\ \bibnamefont
  {Liu}}, \bibinfo {author} {\bibfnamefont {Y.-W.}\ \bibnamefont {Pan}},
  \bibinfo {author} {\bibfnamefont {Z.-W.}\ \bibnamefont {Liu}}, \bibinfo
  {author} {\bibfnamefont {T.-W.}\ \bibnamefont {Wu}}, \bibinfo {author}
  {\bibfnamefont {J.-X.}\ \bibnamefont {Lu}},\ and\ \bibinfo {author}
  {\bibfnamefont {L.-S.}\ \bibnamefont {Geng}},\ }\bibfield  {title} {\bibinfo
  {title} {{Three ways to decipher the nature of exotic hadrons: Multiplets,
  three-body hadronic molecules, and correlation functions}},\ }\href
  {https://doi.org/10.1016/j.physrep.2024.12.001} {\bibfield  {journal}
  {\bibinfo  {journal} {Phys. Rept.}\ }\textbf {\bibinfo {volume} {1108}},\
  \bibinfo {pages} {1} (\bibinfo {year} {2025}{\natexlab{a}})},\ \Eprint
  {https://arxiv.org/abs/2404.06399} {arXiv:2404.06399 [hep-ph]} \BibitemShut
  {NoStop}%
\bibitem [{\citenamefont {Wang}\ \emph {et~al.}(2026)\citenamefont {Wang},
  \citenamefont {Liu},\ and\ \citenamefont {Gao}}]{Wang:2025dur}%
  \BibitemOpen
  \bibfield  {author} {\bibinfo {author} {\bibfnamefont {X.}~\bibnamefont
  {Wang}}, \bibinfo {author} {\bibfnamefont {X.}~\bibnamefont {Liu}},\ and\
  \bibinfo {author} {\bibfnamefont {Y.}~\bibnamefont {Gao}},\ }\bibfield
  {title} {\bibinfo {title} {{Colloquium: Hadron production in open-charm meson
  pairs at e+e- colliders}},\ }\href {https://doi.org/10.1103/2mrp-chly}
  {\bibfield  {journal} {\bibinfo  {journal} {Rev. Mod. Phys.}\ }\textbf
  {\bibinfo {volume} {98}},\ \bibinfo {pages} {021001} (\bibinfo {year}
  {2026})},\ \Eprint {https://arxiv.org/abs/2502.15117} {arXiv:2502.15117
  [hep-ex]} \BibitemShut {NoStop}%
\bibitem [{\citenamefont {Born}\ and\ \citenamefont
  {Oppenheimer}(1927)}]{Born:1927rpw}%
  \BibitemOpen
  \bibfield  {author} {\bibinfo {author} {\bibfnamefont {M.}~\bibnamefont
  {Born}}\ and\ \bibinfo {author} {\bibfnamefont {R.}~\bibnamefont
  {Oppenheimer}},\ }\bibfield  {title} {\bibinfo {title} {{Zur Quantentheorie
  der Molekeln}},\ }\href {https://doi.org/10.1002/andp.19273892002} {\bibfield
   {journal} {\bibinfo  {journal} {Annalen Phys.}\ }\textbf {\bibinfo {volume}
  {389}},\ \bibinfo {pages} {457} (\bibinfo {year} {1927})}\BibitemShut
  {NoStop}%
\bibitem [{\citenamefont {Juge}\ \emph {et~al.}(1999)\citenamefont {Juge},
  \citenamefont {Kuti},\ and\ \citenamefont {Morningstar}}]{Juge:1999ie}%
  \BibitemOpen
  \bibfield  {author} {\bibinfo {author} {\bibfnamefont {K.~J.}\ \bibnamefont
  {Juge}}, \bibinfo {author} {\bibfnamefont {J.}~\bibnamefont {Kuti}},\ and\
  \bibinfo {author} {\bibfnamefont {C.~J.}\ \bibnamefont {Morningstar}},\
  }\bibfield  {title} {\bibinfo {title} {{$Ab initio$ Study of Hybrid $\bar{b}
  g b$ mesons}},\ }\href {https://doi.org/10.1103/PhysRevLett.82.4400}
  {\bibfield  {journal} {\bibinfo  {journal} {Phys. Rev. Lett.}\ }\textbf
  {\bibinfo {volume} {82}},\ \bibinfo {pages} {4400} (\bibinfo {year}
  {1999})},\ \Eprint {https://arxiv.org/abs/hep-ph/9902336}
  {arXiv:hep-ph/9902336} \BibitemShut {NoStop}%
\bibitem [{\citenamefont {Braaten}\ \emph {et~al.}(2014)\citenamefont
  {Braaten}, \citenamefont {Langmack},\ and\ \citenamefont
  {Smith}}]{Braaten:2014qka}%
  \BibitemOpen
  \bibfield  {author} {\bibinfo {author} {\bibfnamefont {E.}~\bibnamefont
  {Braaten}}, \bibinfo {author} {\bibfnamefont {C.}~\bibnamefont {Langmack}},\
  and\ \bibinfo {author} {\bibfnamefont {D.~H.}\ \bibnamefont {Smith}},\
  }\bibfield  {title} {\bibinfo {title} {{Born-Oppenheimer Approximation for
  the XYZ Mesons}},\ }\href {https://doi.org/10.1103/PhysRevD.90.014044}
  {\bibfield  {journal} {\bibinfo  {journal} {Phys. Rev. D}\ }\textbf {\bibinfo
  {volume} {90}},\ \bibinfo {pages} {014044} (\bibinfo {year} {2014})},\
  \Eprint {https://arxiv.org/abs/1402.0438} {arXiv:1402.0438 [hep-ph]}
  \BibitemShut {NoStop}%
\bibitem [{\citenamefont {Bicudo}\ \emph {et~al.}(2015)\citenamefont {Bicudo},
  \citenamefont {Cichy}, \citenamefont {Peters}, \citenamefont {Wagenbach},\
  and\ \citenamefont {Wagner}}]{Bicudo:2015vta}%
  \BibitemOpen
  \bibfield  {author} {\bibinfo {author} {\bibfnamefont {P.}~\bibnamefont
  {Bicudo}}, \bibinfo {author} {\bibfnamefont {K.}~\bibnamefont {Cichy}},
  \bibinfo {author} {\bibfnamefont {A.}~\bibnamefont {Peters}}, \bibinfo
  {author} {\bibfnamefont {B.}~\bibnamefont {Wagenbach}},\ and\ \bibinfo
  {author} {\bibfnamefont {M.}~\bibnamefont {Wagner}},\ }\bibfield  {title}
  {\bibinfo {title} {{Evidence for the existence of $u d \bar{b} \bar{b}$ and
  the non-existence of $s s \bar{b} \bar{b}$ and $c c \bar{b} \bar{b}$
  tetraquarks from lattice QCD}},\ }\href
  {https://doi.org/10.1103/PhysRevD.92.014507} {\bibfield  {journal} {\bibinfo
  {journal} {Phys. Rev. D}\ }\textbf {\bibinfo {volume} {92}},\ \bibinfo
  {pages} {014507} (\bibinfo {year} {2015})},\ \Eprint
  {https://arxiv.org/abs/1505.00613} {arXiv:1505.00613 [hep-lat]} \BibitemShut
  {NoStop}%
\bibitem [{\citenamefont {Bicudo}\ \emph {et~al.}(2017)\citenamefont {Bicudo},
  \citenamefont {Scheunert},\ and\ \citenamefont {Wagner}}]{Bicudo:2016ooe}%
  \BibitemOpen
  \bibfield  {author} {\bibinfo {author} {\bibfnamefont {P.}~\bibnamefont
  {Bicudo}}, \bibinfo {author} {\bibfnamefont {J.}~\bibnamefont {Scheunert}},\
  and\ \bibinfo {author} {\bibfnamefont {M.}~\bibnamefont {Wagner}},\
  }\bibfield  {title} {\bibinfo {title} {{Including heavy spin effects in the
  prediction of a $\bar{b} \bar{b} u d$ tetraquark with lattice QCD
  potentials}},\ }\href {https://doi.org/10.1103/PhysRevD.95.034502} {\bibfield
   {journal} {\bibinfo  {journal} {Phys. Rev. D}\ }\textbf {\bibinfo {volume}
  {95}},\ \bibinfo {pages} {034502} (\bibinfo {year} {2017})},\ \Eprint
  {https://arxiv.org/abs/1612.02758} {arXiv:1612.02758 [hep-lat]} \BibitemShut
  {NoStop}%
\bibitem [{\citenamefont {Bruschini}(2023)}]{Bruschini:2023zkb}%
  \BibitemOpen
  \bibfield  {author} {\bibinfo {author} {\bibfnamefont {R.}~\bibnamefont
  {Bruschini}},\ }\bibfield  {title} {\bibinfo {title} {{Heavy-quark spin
  symmetry breaking in the Born-Oppenheimer approximation}},\ }\href
  {https://doi.org/10.1007/JHEP08(2023)219} {\bibfield  {journal} {\bibinfo
  {journal} {JHEP}\ }\textbf {\bibinfo {volume} {08}},\ \bibinfo {pages}
  {219}},\ \Eprint {https://arxiv.org/abs/2303.17533} {arXiv:2303.17533
  [hep-ph]} \BibitemShut {NoStop}%
\bibitem [{\citenamefont {Berwein}\ \emph {et~al.}(2024)\citenamefont
  {Berwein}, \citenamefont {Brambilla}, \citenamefont {Mohapatra},\ and\
  \citenamefont {Vairo}}]{Berwein:2024ztx}%
  \BibitemOpen
  \bibfield  {author} {\bibinfo {author} {\bibfnamefont {M.}~\bibnamefont
  {Berwein}}, \bibinfo {author} {\bibfnamefont {N.}~\bibnamefont {Brambilla}},
  \bibinfo {author} {\bibfnamefont {A.}~\bibnamefont {Mohapatra}},\ and\
  \bibinfo {author} {\bibfnamefont {A.}~\bibnamefont {Vairo}},\ }\bibfield
  {title} {\bibinfo {title} {{Hybrids, tetraquarks, pentaquarks, doubly heavy
  baryons, and quarkonia in Born-Oppenheimer effective theory}},\ }\href
  {https://doi.org/10.1103/PhysRevD.110.094040} {\bibfield  {journal} {\bibinfo
   {journal} {Phys. Rev. D}\ }\textbf {\bibinfo {volume} {110}},\ \bibinfo
  {pages} {094040} (\bibinfo {year} {2024})},\ \Eprint
  {https://arxiv.org/abs/2408.04719} {arXiv:2408.04719 [hep-ph]} \BibitemShut
  {NoStop}%
\bibitem [{\citenamefont {Braaten}\ and\ \citenamefont
  {Bruschini}(2024)}]{Braaten:2024stn}%
  \BibitemOpen
  \bibfield  {author} {\bibinfo {author} {\bibfnamefont {E.}~\bibnamefont
  {Braaten}}\ and\ \bibinfo {author} {\bibfnamefont {R.}~\bibnamefont
  {Bruschini}},\ }\bibfield  {title} {\bibinfo {title} {{Model-independent
  predictions for decays of hidden-heavy hadrons into pairs of heavy
  hadrons}},\ }\href {https://doi.org/10.1103/PhysRevD.109.094051} {\bibfield
  {journal} {\bibinfo  {journal} {Phys. Rev. D}\ }\textbf {\bibinfo {volume}
  {109}},\ \bibinfo {pages} {094051} (\bibinfo {year} {2024})},\ \Eprint
  {https://arxiv.org/abs/2403.12868} {arXiv:2403.12868 [hep-ph]} \BibitemShut
  {NoStop}%
\bibitem [{\citenamefont {Braaten}\ and\ \citenamefont
  {Bruschini}(2025)}]{Braaten:2024tbm}%
  \BibitemOpen
  \bibfield  {author} {\bibinfo {author} {\bibfnamefont {E.}~\bibnamefont
  {Braaten}}\ and\ \bibinfo {author} {\bibfnamefont {R.}~\bibnamefont
  {Bruschini}},\ }\bibfield  {title} {\bibinfo {title} {{Exotic hidden-heavy
  hadrons and where to find them}},\ }\href
  {https://doi.org/10.1016/j.physletb.2025.139386} {\bibfield  {journal}
  {\bibinfo  {journal} {Phys. Lett. B}\ }\textbf {\bibinfo {volume} {863}},\
  \bibinfo {pages} {139386} (\bibinfo {year} {2025})},\ \Eprint
  {https://arxiv.org/abs/2409.08002} {arXiv:2409.08002 [hep-ph]} \BibitemShut
  {NoStop}%
\bibitem [{\citenamefont {Maiani}\ \emph
  {et~al.}(2019{\natexlab{a}})\citenamefont {Maiani}, \citenamefont {Polosa},\
  and\ \citenamefont {Riquer}}]{Maiani:2019cwl}%
  \BibitemOpen
  \bibfield  {author} {\bibinfo {author} {\bibfnamefont {L.}~\bibnamefont
  {Maiani}}, \bibinfo {author} {\bibfnamefont {A.~D.}\ \bibnamefont {Polosa}},\
  and\ \bibinfo {author} {\bibfnamefont {V.}~\bibnamefont {Riquer}},\
  }\bibfield  {title} {\bibinfo {title} {{Hydrogen bond of QCD}},\ }\href
  {https://doi.org/10.1103/PhysRevD.100.014002} {\bibfield  {journal} {\bibinfo
   {journal} {Phys. Rev. D}\ }\textbf {\bibinfo {volume} {100}},\ \bibinfo
  {pages} {014002} (\bibinfo {year} {2019}{\natexlab{a}})},\ \Eprint
  {https://arxiv.org/abs/1903.10253} {arXiv:1903.10253 [hep-ph]} \BibitemShut
  {NoStop}%
\bibitem [{\citenamefont {Grinstein}\ \emph {et~al.}(2024)\citenamefont
  {Grinstein}, \citenamefont {Maiani},\ and\ \citenamefont
  {Polosa}}]{Grinstein:2024rcu}%
  \BibitemOpen
  \bibfield  {author} {\bibinfo {author} {\bibfnamefont {B.}~\bibnamefont
  {Grinstein}}, \bibinfo {author} {\bibfnamefont {L.}~\bibnamefont {Maiani}},\
  and\ \bibinfo {author} {\bibfnamefont {A.~D.}\ \bibnamefont {Polosa}},\
  }\bibfield  {title} {\bibinfo {title} {{Radiative decays of X(3872)
  discriminate between the molecular and compact interpretations}},\ }\href
  {https://doi.org/10.1103/PhysRevD.109.074009} {\bibfield  {journal} {\bibinfo
   {journal} {Phys. Rev. D}\ }\textbf {\bibinfo {volume} {109}},\ \bibinfo
  {pages} {074009} (\bibinfo {year} {2024})},\ \Eprint
  {https://arxiv.org/abs/2401.11623} {arXiv:2401.11623 [hep-ph]} \BibitemShut
  {NoStop}%
\bibitem [{\citenamefont {Liu}\ \emph {et~al.}(2025{\natexlab{b}})\citenamefont
  {Liu}, \citenamefont {Xiao},\ and\ \citenamefont {Guo}}]{Liu:2025jyn}%
  \BibitemOpen
  \bibfield  {author} {\bibinfo {author} {\bibfnamefont {L.}~\bibnamefont
  {Liu}}, \bibinfo {author} {\bibfnamefont {Y.}~\bibnamefont {Xiao}},\ and\
  \bibinfo {author} {\bibfnamefont {T.}~\bibnamefont {Guo}},\ }\bibfield
  {title} {\bibinfo {title} {{Hydrogenlike structures in the strong
  interaction}},\ }\href {https://doi.org/10.1103/77hs-2gy4} {\bibfield
  {journal} {\bibinfo  {journal} {Phys. Rev. D}\ }\textbf {\bibinfo {volume}
  {112}},\ \bibinfo {pages} {054021} (\bibinfo {year} {2025}{\natexlab{b}})},\
  \Eprint {https://arxiv.org/abs/2505.22177} {arXiv:2505.22177 [hep-ph]}
  \BibitemShut {NoStop}%
\bibitem [{\citenamefont {Germani}\ \emph {et~al.}(2025)\citenamefont
  {Germani}, \citenamefont {Grinstein},\ and\ \citenamefont
  {Polosa}}]{Germani:2025mos}%
  \BibitemOpen
  \bibfield  {author} {\bibinfo {author} {\bibfnamefont {D.}~\bibnamefont
  {Germani}}, \bibinfo {author} {\bibfnamefont {B.}~\bibnamefont {Grinstein}},\
  and\ \bibinfo {author} {\bibfnamefont {A.~D.}\ \bibnamefont {Polosa}},\
  }\bibfield  {title} {\bibinfo {title} {{Tetraquarks in the Born-Oppenheimer
  approximation}},\ }\href {https://doi.org/10.1007/JHEP04(2025)004} {\bibfield
   {journal} {\bibinfo  {journal} {JHEP}\ }\textbf {\bibinfo {volume} {04}},\
  \bibinfo {pages} {004}},\ \Eprint {https://arxiv.org/abs/2501.13249}
  {arXiv:2501.13249 [hep-ph]} \BibitemShut {NoStop}%
\bibitem [{\citenamefont {Kang}\ \emph {et~al.}(2025)\citenamefont {Kang},
  \citenamefont {Xia},\ and\ \citenamefont {Guo}}]{Kang:2025xqm}%
  \BibitemOpen
  \bibfield  {author} {\bibinfo {author} {\bibfnamefont {B.}~\bibnamefont
  {Kang}}, \bibinfo {author} {\bibfnamefont {X.}~\bibnamefont {Xia}},\ and\
  \bibinfo {author} {\bibfnamefont {T.}~\bibnamefont {Guo}},\ }\bibfield
  {title} {\bibinfo {title} {{Hidden heavy flavor tetraquarks in the
  Born-Oppenheimer approximation}},\ }\href {https://doi.org/10.1103/xnbg-wklj}
  {\bibfield  {journal} {\bibinfo  {journal} {Phys. Rev. D}\ }\textbf {\bibinfo
  {volume} {111}},\ \bibinfo {pages} {114016} (\bibinfo {year} {2025})},\
  \Eprint {https://arxiv.org/abs/2503.10173} {arXiv:2503.10173 [hep-ph]}
  \BibitemShut {NoStop}%
\bibitem [{\citenamefont {Pauling}(1928)}]{pauling1928application}%
  \BibitemOpen
  \bibfield  {author} {\bibinfo {author} {\bibfnamefont {L.}~\bibnamefont
  {Pauling}},\ }\bibfield  {title} {\bibinfo {title} {The application of the
  quantum mechanics to the structure of the hydrogen molecule and hydrogen
  molecule-ion and to related problems.},\ }\href@noop {} {\bibfield  {journal}
  {\bibinfo  {journal} {Chemical Reviews}\ }\textbf {\bibinfo {volume} {5}},\
  \bibinfo {pages} {173} (\bibinfo {year} {1928})}\BibitemShut {NoStop}%
\bibitem [{\citenamefont {Szabo}\ and\ \citenamefont
  {Ostlund}(1996)}]{szabo2012modern}%
  \BibitemOpen
  \bibfield  {author} {\bibinfo {author} {\bibfnamefont {A.}~\bibnamefont
  {Szabo}}\ and\ \bibinfo {author} {\bibfnamefont {N.~S.}\ \bibnamefont
  {Ostlund}},\ }\href@noop {} {\emph {\bibinfo {title} {Modern quantum
  chemistry: introduction to advanced electronic structure theory}}}\ (\bibinfo
   {publisher} {Dover Publications},\ \bibinfo {year} {1996})\BibitemShut
  {NoStop}%
\bibitem [{\citenamefont {Hiyama}\ \emph {et~al.}(2003)\citenamefont {Hiyama},
  \citenamefont {Kino},\ and\ \citenamefont {Kamimura}}]{Hiyama:2003cu}%
  \BibitemOpen
  \bibfield  {author} {\bibinfo {author} {\bibfnamefont {E.}~\bibnamefont
  {Hiyama}}, \bibinfo {author} {\bibfnamefont {Y.}~\bibnamefont {Kino}},\ and\
  \bibinfo {author} {\bibfnamefont {M.}~\bibnamefont {Kamimura}},\ }\bibfield
  {title} {\bibinfo {title} {{Gaussian expansion method for few-body
  systems}},\ }\href {https://doi.org/10.1016/S0146-6410(03)90015-9} {\bibfield
   {journal} {\bibinfo  {journal} {Prog. Part. Nucl. Phys.}\ }\textbf {\bibinfo
  {volume} {51}},\ \bibinfo {pages} {223} (\bibinfo {year} {2003})}\BibitemShut
  {NoStop}%
\bibitem [{\citenamefont {Hiyama}(2012)}]{Hiyama:2012sma}%
  \BibitemOpen
  \bibfield  {author} {\bibinfo {author} {\bibfnamefont {E.}~\bibnamefont
  {Hiyama}},\ }\bibfield  {title} {\bibinfo {title} {{Gaussian expansion method
  for few-body systems and its applications to atomic and nuclear physics}},\
  }\href {https://doi.org/10.1093/ptep/pts015} {\bibfield  {journal} {\bibinfo
  {journal} {PTEP}\ }\textbf {\bibinfo {volume} {2012}},\ \bibinfo {pages}
  {01A204} (\bibinfo {year} {2012})}\BibitemShut {NoStop}%
\bibitem [{\citenamefont {Luo}\ and\ \citenamefont
  {Liu}(2023{\natexlab{a}})}]{Luo:2023sne}%
  \BibitemOpen
  \bibfield  {author} {\bibinfo {author} {\bibfnamefont {S.-Q.}\ \bibnamefont
  {Luo}}\ and\ \bibinfo {author} {\bibfnamefont {X.}~\bibnamefont {Liu}},\
  }\bibfield  {title} {\bibinfo {title} {{Investigating the spectroscopy
  behavior of undetected $1F$-wave charmed baryons}},\ }\href
  {https://doi.org/10.1103/PhysRevD.108.034002} {\bibfield  {journal} {\bibinfo
   {journal} {Phys. Rev. D}\ }\textbf {\bibinfo {volume} {108}},\ \bibinfo
  {pages} {034002} (\bibinfo {year} {2023}{\natexlab{a}})},\ \Eprint
  {https://arxiv.org/abs/2306.04588} {arXiv:2306.04588 [hep-ph]} \BibitemShut
  {NoStop}%
\bibitem [{\citenamefont {Man}\ \emph {et~al.}(2024)\citenamefont {Man},
  \citenamefont {Shu}, \citenamefont {Liu},\ and\ \citenamefont
  {Chen}}]{Man:2024mvl}%
  \BibitemOpen
  \bibfield  {author} {\bibinfo {author} {\bibfnamefont {Z.-L.}\ \bibnamefont
  {Man}}, \bibinfo {author} {\bibfnamefont {C.-R.}\ \bibnamefont {Shu}},
  \bibinfo {author} {\bibfnamefont {Y.-R.}\ \bibnamefont {Liu}},\ and\ \bibinfo
  {author} {\bibfnamefont {H.}~\bibnamefont {Chen}},\ }\bibfield  {title}
  {\bibinfo {title} {{Charmonium states in a coupled-channel model}},\ }\href
  {https://doi.org/10.1140/epjc/s10052-024-13132-7} {\bibfield  {journal}
  {\bibinfo  {journal} {Eur. Phys. J. C}\ }\textbf {\bibinfo {volume} {84}},\
  \bibinfo {pages} {810} (\bibinfo {year} {2024})},\ \Eprint
  {https://arxiv.org/abs/2402.02765} {arXiv:2402.02765 [hep-ph]} \BibitemShut
  {NoStop}%
\bibitem [{\citenamefont {Zhou}\ \emph {et~al.}(2025)\citenamefont {Zhou},
  \citenamefont {Luo},\ and\ \citenamefont {Liu}}]{Zhou:2025fpp}%
  \BibitemOpen
  \bibfield  {author} {\bibinfo {author} {\bibfnamefont {H.}~\bibnamefont
  {Zhou}}, \bibinfo {author} {\bibfnamefont {S.-Q.}\ \bibnamefont {Luo}},\ and\
  \bibinfo {author} {\bibfnamefont {X.}~\bibnamefont {Liu}},\ }\bibfield
  {title} {\bibinfo {title} {{Triply heavy baryon spectroscopy revisited}},\
  }\href {https://doi.org/10.1103/jhr1-ccsw} {\bibfield  {journal} {\bibinfo
  {journal} {Phys. Rev. D}\ }\textbf {\bibinfo {volume} {112}},\ \bibinfo
  {pages} {074007} (\bibinfo {year} {2025})},\ \Eprint
  {https://arxiv.org/abs/2507.10243} {arXiv:2507.10243 [hep-ph]} \BibitemShut
  {NoStop}%
\bibitem [{\citenamefont {Luo}\ \emph {et~al.}(2025)\citenamefont {Luo},
  \citenamefont {Huang},\ and\ \citenamefont {Liu}}]{Luo:2025psq}%
  \BibitemOpen
  \bibfield  {author} {\bibinfo {author} {\bibfnamefont {S.-Q.}\ \bibnamefont
  {Luo}}, \bibinfo {author} {\bibfnamefont {Q.}~\bibnamefont {Huang}},\ and\
  \bibinfo {author} {\bibfnamefont {X.}~\bibnamefont {Liu}},\ }\bibfield
  {title} {\bibinfo {title} {{The quest for topped hadrons}},\ }\href@noop {}
  {\  (\bibinfo {year} {2025})},\ \Eprint {https://arxiv.org/abs/2508.17646}
  {arXiv:2508.17646 [hep-ph]} \BibitemShut {NoStop}%
\bibitem [{\citenamefont {Zhang}\ and\ \citenamefont
  {Luo}(2025)}]{Zhang:2025vtc}%
  \BibitemOpen
  \bibfield  {author} {\bibinfo {author} {\bibfnamefont {Z.-L.}\ \bibnamefont
  {Zhang}}\ and\ \bibinfo {author} {\bibfnamefont {S.-Q.}\ \bibnamefont
  {Luo}},\ }\bibfield  {title} {\bibinfo {title} {{Spectroscopic properties of
  1F-wave singly bottom baryons}},\ }\href {https://doi.org/10.1103/y6gk-2dcw}
  {\bibfield  {journal} {\bibinfo  {journal} {Phys. Rev. D}\ }\textbf {\bibinfo
  {volume} {112}},\ \bibinfo {pages} {074020} (\bibinfo {year} {2025})},\
  \Eprint {https://arxiv.org/abs/2504.17507} {arXiv:2504.17507 [hep-ph]}
  \BibitemShut {NoStop}%
\bibitem [{\citenamefont {Luo}\ and\ \citenamefont {Liu}(2025)}]{Luo:2025cqs}%
  \BibitemOpen
  \bibfield  {author} {\bibinfo {author} {\bibfnamefont {S.-Q.}\ \bibnamefont
  {Luo}}\ and\ \bibinfo {author} {\bibfnamefont {X.}~\bibnamefont {Liu}},\
  }\bibfield  {title} {\bibinfo {title} {{Identifying triple-strangeness
  {\ensuremath{\Omega}} hyperons in light of recent experimental results}},\
  }\href {https://doi.org/10.1103/18md-j4bf} {\bibfield  {journal} {\bibinfo
  {journal} {Phys. Rev. D}\ }\textbf {\bibinfo {volume} {112}},\ \bibinfo
  {pages} {014047} (\bibinfo {year} {2025})},\ \Eprint
  {https://arxiv.org/abs/2504.14648} {arXiv:2504.14648 [hep-ph]} \BibitemShut
  {NoStop}%
\bibitem [{\citenamefont {An}\ \emph {et~al.}(2025)\citenamefont {An},
  \citenamefont {Luo},\ and\ \citenamefont {Liu}}]{An:2025rjv}%
  \BibitemOpen
  \bibfield  {author} {\bibinfo {author} {\bibfnamefont {H.-T.}\ \bibnamefont
  {An}}, \bibinfo {author} {\bibfnamefont {S.-Q.}\ \bibnamefont {Luo}},\ and\
  \bibinfo {author} {\bibfnamefont {X.}~\bibnamefont {Liu}},\ }\bibfield
  {title} {\bibinfo {title} {{Doubly charmed hexaquarks in the diquark
  picture}},\ }\href {https://doi.org/10.1103/l4qq-pbr9} {\bibfield  {journal}
  {\bibinfo  {journal} {Phys. Rev. D}\ }\textbf {\bibinfo {volume} {112}},\
  \bibinfo {pages} {054041} (\bibinfo {year} {2025})},\ \Eprint
  {https://arxiv.org/abs/2504.06107} {arXiv:2504.06107 [hep-ph]} \BibitemShut
  {NoStop}%
\bibitem [{\citenamefont {Meng}\ \emph {et~al.}(2023)\citenamefont {Meng},
  \citenamefont {Chen}, \citenamefont {Ma},\ and\ \citenamefont
  {Zhu}}]{Meng:2023jqk}%
  \BibitemOpen
  \bibfield  {author} {\bibinfo {author} {\bibfnamefont {L.}~\bibnamefont
  {Meng}}, \bibinfo {author} {\bibfnamefont {Y.-K.}\ \bibnamefont {Chen}},
  \bibinfo {author} {\bibfnamefont {Y.}~\bibnamefont {Ma}},\ and\ \bibinfo
  {author} {\bibfnamefont {S.-L.}\ \bibnamefont {Zhu}},\ }\bibfield  {title}
  {\bibinfo {title} {{Tetraquark bound states in constituent quark models:
  Benchmark test calculations}},\ }\href
  {https://doi.org/10.1103/PhysRevD.108.114016} {\bibfield  {journal} {\bibinfo
   {journal} {Phys. Rev. D}\ }\textbf {\bibinfo {volume} {108}},\ \bibinfo
  {pages} {114016} (\bibinfo {year} {2023})},\ \Eprint
  {https://arxiv.org/abs/2310.13354} {arXiv:2310.13354 [hep-ph]} \BibitemShut
  {NoStop}%
\bibitem [{\citenamefont {Wu}\ \emph {et~al.}(2024{\natexlab{a}})\citenamefont
  {Wu}, \citenamefont {Ma}, \citenamefont {Chen}, \citenamefont {Meng},\ and\
  \citenamefont {Zhu}}]{Wu:2024hrv}%
  \BibitemOpen
  \bibfield  {author} {\bibinfo {author} {\bibfnamefont {W.-L.}\ \bibnamefont
  {Wu}}, \bibinfo {author} {\bibfnamefont {Y.}~\bibnamefont {Ma}}, \bibinfo
  {author} {\bibfnamefont {Y.-K.}\ \bibnamefont {Chen}}, \bibinfo {author}
  {\bibfnamefont {L.}~\bibnamefont {Meng}},\ and\ \bibinfo {author}
  {\bibfnamefont {S.-L.}\ \bibnamefont {Zhu}},\ }\bibfield  {title} {\bibinfo
  {title} {{Fully heavy tetraquark resonant states with different flavors}},\
  }\href {https://doi.org/10.1103/PhysRevD.110.034030} {\bibfield  {journal}
  {\bibinfo  {journal} {Phys. Rev. D}\ }\textbf {\bibinfo {volume} {110}},\
  \bibinfo {pages} {034030} (\bibinfo {year} {2024}{\natexlab{a}})},\ \Eprint
  {https://arxiv.org/abs/2406.17824} {arXiv:2406.17824 [hep-ph]} \BibitemShut
  {NoStop}%
\bibitem [{\citenamefont {Wu}\ \emph {et~al.}(2024{\natexlab{b}})\citenamefont
  {Wu}, \citenamefont {Chen}, \citenamefont {Meng},\ and\ \citenamefont
  {Zhu}}]{Wu:2024euj}%
  \BibitemOpen
  \bibfield  {author} {\bibinfo {author} {\bibfnamefont {W.-L.}\ \bibnamefont
  {Wu}}, \bibinfo {author} {\bibfnamefont {Y.-K.}\ \bibnamefont {Chen}},
  \bibinfo {author} {\bibfnamefont {L.}~\bibnamefont {Meng}},\ and\ \bibinfo
  {author} {\bibfnamefont {S.-L.}\ \bibnamefont {Zhu}},\ }\bibfield  {title}
  {\bibinfo {title} {{Benchmark calculations of fully heavy compact and
  molecular tetraquark states}},\ }\href
  {https://doi.org/10.1103/PhysRevD.109.054034} {\bibfield  {journal} {\bibinfo
   {journal} {Phys. Rev. D}\ }\textbf {\bibinfo {volume} {109}},\ \bibinfo
  {pages} {054034} (\bibinfo {year} {2024}{\natexlab{b}})},\ \Eprint
  {https://arxiv.org/abs/2401.14899} {arXiv:2401.14899 [hep-ph]} \BibitemShut
  {NoStop}%
\bibitem [{\citenamefont {Wu}\ \emph {et~al.}(2024{\natexlab{c}})\citenamefont
  {Wu}, \citenamefont {Ma}, \citenamefont {Chen}, \citenamefont {Meng},\ and\
  \citenamefont {Zhu}}]{Wu:2024zbx}%
  \BibitemOpen
  \bibfield  {author} {\bibinfo {author} {\bibfnamefont {W.-L.}\ \bibnamefont
  {Wu}}, \bibinfo {author} {\bibfnamefont {Y.}~\bibnamefont {Ma}}, \bibinfo
  {author} {\bibfnamefont {Y.-K.}\ \bibnamefont {Chen}}, \bibinfo {author}
  {\bibfnamefont {L.}~\bibnamefont {Meng}},\ and\ \bibinfo {author}
  {\bibfnamefont {S.-L.}\ \bibnamefont {Zhu}},\ }\bibfield  {title} {\bibinfo
  {title} {{Doubly heavy tetraquark bound and resonant states}},\ }\href
  {https://doi.org/10.1103/PhysRevD.110.094041} {\bibfield  {journal} {\bibinfo
   {journal} {Phys. Rev. D}\ }\textbf {\bibinfo {volume} {110}},\ \bibinfo
  {pages} {094041} (\bibinfo {year} {2024}{\natexlab{c}})},\ \Eprint
  {https://arxiv.org/abs/2409.03373} {arXiv:2409.03373 [hep-ph]} \BibitemShut
  {NoStop}%
\bibitem [{\citenamefont {Luo}\ and\ \citenamefont
  {Liu}(2023{\natexlab{b}})}]{Luo:2023sra}%
  \BibitemOpen
  \bibfield  {author} {\bibinfo {author} {\bibfnamefont {S.-Q.}\ \bibnamefont
  {Luo}}\ and\ \bibinfo {author} {\bibfnamefont {X.}~\bibnamefont {Liu}},\
  }\bibfield  {title} {\bibinfo {title} {{Newly observed
  {\ensuremath{\Omega}}c(3327): A good candidate for a D-wave charmed
  baryon}},\ }\href {https://doi.org/10.1103/PhysRevD.107.074041} {\bibfield
  {journal} {\bibinfo  {journal} {Phys. Rev. D}\ }\textbf {\bibinfo {volume}
  {107}},\ \bibinfo {pages} {074041} (\bibinfo {year} {2023}{\natexlab{b}})},\
  \Eprint {https://arxiv.org/abs/2303.04022} {arXiv:2303.04022 [hep-ph]}
  \BibitemShut {NoStop}%
\bibitem [{\citenamefont {Peng}\ \emph {et~al.}(2024)\citenamefont {Peng},
  \citenamefont {Luo},\ and\ \citenamefont {Liu}}]{Peng:2024pyl}%
  \BibitemOpen
  \bibfield  {author} {\bibinfo {author} {\bibfnamefont {Y.-X.}\ \bibnamefont
  {Peng}}, \bibinfo {author} {\bibfnamefont {S.-Q.}\ \bibnamefont {Luo}},\ and\
  \bibinfo {author} {\bibfnamefont {X.}~\bibnamefont {Liu}},\ }\bibfield
  {title} {\bibinfo {title} {{Refining radiative decay studies in singly heavy
  baryons}},\ }\href {https://doi.org/10.1103/PhysRevD.110.074034} {\bibfield
  {journal} {\bibinfo  {journal} {Phys. Rev. D}\ }\textbf {\bibinfo {volume}
  {110}},\ \bibinfo {pages} {074034} (\bibinfo {year} {2024})},\ \Eprint
  {https://arxiv.org/abs/2405.12812} {arXiv:2405.12812 [hep-ph]} \BibitemShut
  {NoStop}%
\bibitem [{\citenamefont {Luo}\ \emph {et~al.}(2021)\citenamefont {Luo},
  \citenamefont {Chen}, \citenamefont {Liu},\ and\ \citenamefont
  {Matsuki}}]{Luo:2021dvj}%
  \BibitemOpen
  \bibfield  {author} {\bibinfo {author} {\bibfnamefont {S.-Q.}\ \bibnamefont
  {Luo}}, \bibinfo {author} {\bibfnamefont {B.}~\bibnamefont {Chen}}, \bibinfo
  {author} {\bibfnamefont {X.}~\bibnamefont {Liu}},\ and\ \bibinfo {author}
  {\bibfnamefont {T.}~\bibnamefont {Matsuki}},\ }\bibfield  {title} {\bibinfo
  {title} {{Predicting a new resonance as charmed-strange baryonic analog of
  $D^*_{s0}$(2317)}},\ }\href {https://doi.org/10.1103/PhysRevD.103.074027}
  {\bibfield  {journal} {\bibinfo  {journal} {Phys. Rev. D}\ }\textbf {\bibinfo
  {volume} {103}},\ \bibinfo {pages} {074027} (\bibinfo {year} {2021})},\
  \Eprint {https://arxiv.org/abs/2102.00679} {arXiv:2102.00679 [hep-ph]}
  \BibitemShut {NoStop}%
\bibitem [{\citenamefont {Luo}\ \emph {et~al.}(2023)\citenamefont {Luo},
  \citenamefont {Liu},\ and\ \citenamefont {Liu}}]{Luo:2023hnp}%
  \BibitemOpen
  \bibfield  {author} {\bibinfo {author} {\bibfnamefont {S.-Q.}\ \bibnamefont
  {Luo}}, \bibinfo {author} {\bibfnamefont {Z.-W.}\ \bibnamefont {Liu}},\ and\
  \bibinfo {author} {\bibfnamefont {X.}~\bibnamefont {Liu}},\ }\bibfield
  {title} {\bibinfo {title} {{New type of hydrogenlike charm-pion or charm-kaon
  matter}},\ }\href {https://doi.org/10.1103/PhysRevD.107.054022} {\bibfield
  {journal} {\bibinfo  {journal} {Phys. Rev. D}\ }\textbf {\bibinfo {volume}
  {107}},\ \bibinfo {pages} {054022} (\bibinfo {year} {2023})},\ \Eprint
  {https://arxiv.org/abs/2302.13202} {arXiv:2302.13202 [hep-ph]} \BibitemShut
  {NoStop}%
\bibitem [{\citenamefont {Weinberg}(2015)}]{Weinberg_2015}%
  \BibitemOpen
  \bibfield  {author} {\bibinfo {author} {\bibfnamefont {S.}~\bibnamefont
  {Weinberg}},\ }\href@noop {} {\emph {\bibinfo {title} {{Lectures on Quantum
  Mechanics}}}}\ (\bibinfo  {publisher} {Cambridge University Press},\ \bibinfo
  {year} {2015})\BibitemShut {NoStop}%
\bibitem [{\citenamefont {Bratsev}(1965)}]{Bratsev:1965}%
  \BibitemOpen
  \bibfield  {author} {\bibinfo {author} {\bibfnamefont {V.~F.}\ \bibnamefont
  {Bratsev}},\ }\bibfield  {title} {\bibinfo {title} {{The ground state energy
  of a molecule in adiabatic approximation}},\ }\href
  {http://mi.mathnet.ru/dan30611} {\bibfield  {journal} {\bibinfo  {journal}
  {Dokl. Akad. Nauk SSSR}\ }\textbf {\bibinfo {volume} {160}},\ \bibinfo
  {pages} {570} (\bibinfo {year} {1965})},\ \bibinfo {note} {[Proc. Acad. Sci.
  USSR]}\BibitemShut {NoStop}%
\bibitem [{\citenamefont {Das}\ \emph {et~al.}(1993)\citenamefont {Das},
  \citenamefont {Coelho},\ and\ \citenamefont {Brito}}]{Das:1993zz}%
  \BibitemOpen
  \bibfield  {author} {\bibinfo {author} {\bibfnamefont {T.~K.}\ \bibnamefont
  {Das}}, \bibinfo {author} {\bibfnamefont {H.~T.}\ \bibnamefont {Coelho}},\
  and\ \bibinfo {author} {\bibfnamefont {V.~P.}\ \bibnamefont {Brito}},\
  }\bibfield  {title} {\bibinfo {title} {{Comparison of Born-Oppenheimer and
  hyperspherical adiabatic approximations in the trinucleon problem}},\ }\href
  {https://doi.org/10.1103/PhysRevC.48.2201} {\bibfield  {journal} {\bibinfo
  {journal} {Phys. Rev. C}\ }\textbf {\bibinfo {volume} {48}},\ \bibinfo
  {pages} {2201} (\bibinfo {year} {1993})}\BibitemShut {NoStop}%
\bibitem [{\citenamefont {Maiani}\ \emph
  {et~al.}(2019{\natexlab{b}})\citenamefont {Maiani}, \citenamefont {Polosa},\
  and\ \citenamefont {Riquer}}]{Maiani:2019lpu}%
  \BibitemOpen
  \bibfield  {author} {\bibinfo {author} {\bibfnamefont {L.}~\bibnamefont
  {Maiani}}, \bibinfo {author} {\bibfnamefont {A.~D.}\ \bibnamefont {Polosa}},\
  and\ \bibinfo {author} {\bibfnamefont {V.}~\bibnamefont {Riquer}},\
  }\bibfield  {title} {\bibinfo {title} {{Hydrogen bond of QCD in doubly heavy
  baryons and tetraquarks}},\ }\href
  {https://doi.org/10.1103/PhysRevD.100.074002} {\bibfield  {journal} {\bibinfo
   {journal} {Phys. Rev. D}\ }\textbf {\bibinfo {volume} {100}},\ \bibinfo
  {pages} {074002} (\bibinfo {year} {2019}{\natexlab{b}})},\ \Eprint
  {https://arxiv.org/abs/1908.03244} {arXiv:1908.03244 [hep-ph]} \BibitemShut
  {NoStop}%
\bibitem [{\citenamefont {Li}\ \emph {et~al.}(2024{\natexlab{a}})\citenamefont
  {Li}, \citenamefont {Liu}, \citenamefont {Man}, \citenamefont {Si},\ and\
  \citenamefont {Wu}}]{Li:2023wug}%
  \BibitemOpen
  \bibfield  {author} {\bibinfo {author} {\bibfnamefont {S.-Y.}\ \bibnamefont
  {Li}}, \bibinfo {author} {\bibfnamefont {Y.-R.}\ \bibnamefont {Liu}},
  \bibinfo {author} {\bibfnamefont {Z.-L.}\ \bibnamefont {Man}}, \bibinfo
  {author} {\bibfnamefont {Z.-G.}\ \bibnamefont {Si}},\ and\ \bibinfo {author}
  {\bibfnamefont {J.}~\bibnamefont {Wu}},\ }\bibfield  {title} {\bibinfo
  {title} {{Doubly heavy tetraquark states in a mass splitting model}},\ }\href
  {https://doi.org/10.1103/PhysRevD.110.094044} {\bibfield  {journal} {\bibinfo
   {journal} {Phys. Rev. D}\ }\textbf {\bibinfo {volume} {110}},\ \bibinfo
  {pages} {094044} (\bibinfo {year} {2024}{\natexlab{a}})},\ \Eprint
  {https://arxiv.org/abs/2401.00115} {arXiv:2401.00115 [hep-ph]} \BibitemShut
  {NoStop}%
\bibitem [{\citenamefont {Li}\ \emph {et~al.}(2024{\natexlab{b}})\citenamefont
  {Li}, \citenamefont {Liu}, \citenamefont {Man}, \citenamefont {Si},\ and\
  \citenamefont {Wu}}]{Li:2023wxm}%
  \BibitemOpen
  \bibfield  {author} {\bibinfo {author} {\bibfnamefont {S.-Y.}\ \bibnamefont
  {Li}}, \bibinfo {author} {\bibfnamefont {Y.-R.}\ \bibnamefont {Liu}},
  \bibinfo {author} {\bibfnamefont {Z.-L.}\ \bibnamefont {Man}}, \bibinfo
  {author} {\bibfnamefont {Z.-G.}\ \bibnamefont {Si}},\ and\ \bibinfo {author}
  {\bibfnamefont {J.}~\bibnamefont {Wu}},\ }\bibfield  {title} {\bibinfo
  {title} {{X(3960), X $_{0}$(4140), and other compact states*}},\ }\href
  {https://doi.org/10.1088/1674-1137/ad34c4} {\bibfield  {journal} {\bibinfo
  {journal} {Chin. Phys. C}\ }\textbf {\bibinfo {volume} {48}},\ \bibinfo
  {pages} {063109} (\bibinfo {year} {2024}{\natexlab{b}})},\ \Eprint
  {https://arxiv.org/abs/2308.06768} {arXiv:2308.06768 [hep-ph]} \BibitemShut
  {NoStop}%
\bibitem [{\citenamefont {Li}\ \emph {et~al.}(2023)\citenamefont {Li},
  \citenamefont {Liu}, \citenamefont {Man}, \citenamefont {Si},\ and\
  \citenamefont {Wu}}]{Li:2023aui}%
  \BibitemOpen
  \bibfield  {author} {\bibinfo {author} {\bibfnamefont {S.-Y.}\ \bibnamefont
  {Li}}, \bibinfo {author} {\bibfnamefont {Y.-R.}\ \bibnamefont {Liu}},
  \bibinfo {author} {\bibfnamefont {Z.-L.}\ \bibnamefont {Man}}, \bibinfo
  {author} {\bibfnamefont {Z.-G.}\ \bibnamefont {Si}},\ and\ \bibinfo {author}
  {\bibfnamefont {J.}~\bibnamefont {Wu}},\ }\bibfield  {title} {\bibinfo
  {title} {{Hidden-charm pentaquark states in a mass splitting model}},\ }\href
  {https://doi.org/10.1103/PhysRevD.108.056015} {\bibfield  {journal} {\bibinfo
   {journal} {Phys. Rev. D}\ }\textbf {\bibinfo {volume} {108}},\ \bibinfo
  {pages} {056015} (\bibinfo {year} {2023})},\ \Eprint
  {https://arxiv.org/abs/2307.00539} {arXiv:2307.00539 [hep-ph]} \BibitemShut
  {NoStop}%
\bibitem [{\citenamefont {Li}\ \emph {et~al.}(2025)\citenamefont {Li},
  \citenamefont {Liu}, \citenamefont {Man}, \citenamefont {Shu}, \citenamefont
  {Si},\ and\ \citenamefont {Wu}}]{Li:2025fmf}%
  \BibitemOpen
  \bibfield  {author} {\bibinfo {author} {\bibfnamefont {S.-Y.}\ \bibnamefont
  {Li}}, \bibinfo {author} {\bibfnamefont {Y.-R.}\ \bibnamefont {Liu}},
  \bibinfo {author} {\bibfnamefont {Z.-L.}\ \bibnamefont {Man}}, \bibinfo
  {author} {\bibfnamefont {C.-R.}\ \bibnamefont {Shu}}, \bibinfo {author}
  {\bibfnamefont {Z.-G.}\ \bibnamefont {Si}},\ and\ \bibinfo {author}
  {\bibfnamefont {J.}~\bibnamefont {Wu}},\ }\bibfield  {title} {\bibinfo
  {title} {{Triply Heavy Tetraquark States in a Mass-Splitting Model}},\ }\href
  {https://doi.org/10.3390/sym17020170} {\bibfield  {journal} {\bibinfo
  {journal} {Symmetry}\ }\textbf {\bibinfo {volume} {17}},\ \bibinfo {pages}
  {170} (\bibinfo {year} {2025})},\ \Eprint {https://arxiv.org/abs/2501.16105}
  {arXiv:2501.16105 [hep-ph]} \BibitemShut {NoStop}%
\bibitem [{\citenamefont {Richardson}(1979)}]{Richardson:1978bt}%
  \BibitemOpen
  \bibfield  {author} {\bibinfo {author} {\bibfnamefont {J.~L.}\ \bibnamefont
  {Richardson}},\ }\bibfield  {title} {\bibinfo {title} {{The Heavy Quark
  Potential and the Upsilon, $J/\psi$ Systems}},\ }\href
  {https://doi.org/10.1016/0370-2693(79)90753-6} {\bibfield  {journal}
  {\bibinfo  {journal} {Phys. Lett. B}\ }\textbf {\bibinfo {volume} {82}},\
  \bibinfo {pages} {272} (\bibinfo {year} {1979})}\BibitemShut {NoStop}%
\bibitem [{\citenamefont {Tang}\ \emph {et~al.}(1995)\citenamefont {Tang},
  \citenamefont {Liu},\ and\ \citenamefont {Chao}}]{Tang:1995iy}%
  \BibitemOpen
  \bibfield  {author} {\bibinfo {author} {\bibfnamefont {J.}~\bibnamefont
  {Tang}}, \bibinfo {author} {\bibfnamefont {J.-H.}\ \bibnamefont {Liu}},\ and\
  \bibinfo {author} {\bibfnamefont {K.-T.}\ \bibnamefont {Chao}},\ }\bibfield
  {title} {\bibinfo {title} {{Hadronic matrix elements and radiative $B \to
  K^*\gamma$ decay}},\ }\href {https://doi.org/10.1103/PhysRevD.51.3501}
  {\bibfield  {journal} {\bibinfo  {journal} {Phys. Rev. D}\ }\textbf {\bibinfo
  {volume} {51}},\ \bibinfo {pages} {3501} (\bibinfo {year} {1995})},\ \Eprint
  {https://arxiv.org/abs/hep-ph/9502411} {arXiv:hep-ph/9502411} \BibitemShut
  {NoStop}%
\bibitem [{\citenamefont {Wang}\ \emph {et~al.}(2025)\citenamefont {Wang},
  \citenamefont {Feng},\ and\ \citenamefont {Wang}}]{Wang:2024hzd}%
  \BibitemOpen
  \bibfield  {author} {\bibinfo {author} {\bibfnamefont {G.-L.}\ \bibnamefont
  {Wang}}, \bibinfo {author} {\bibfnamefont {T.-F.}\ \bibnamefont {Feng}},\
  and\ \bibinfo {author} {\bibfnamefont {Y.-Q.}\ \bibnamefont {Wang}},\
  }\bibfield  {title} {\bibinfo {title} {{Mass spectra and wave functions of
  toponia}},\ }\href {https://doi.org/10.1103/PhysRevD.111.096016} {\bibfield
  {journal} {\bibinfo  {journal} {Phys. Rev. D}\ }\textbf {\bibinfo {volume}
  {111}},\ \bibinfo {pages} {096016} (\bibinfo {year} {2025})},\ \Eprint
  {https://arxiv.org/abs/2411.17955} {arXiv:2411.17955 [hep-ph]} \BibitemShut
  {NoStop}%
\bibitem [{\citenamefont {Kawanai}\ and\ \citenamefont
  {Sasaki}(2012)}]{Kawanai:2011jt}%
  \BibitemOpen
  \bibfield  {author} {\bibinfo {author} {\bibfnamefont {T.}~\bibnamefont
  {Kawanai}}\ and\ \bibinfo {author} {\bibfnamefont {S.}~\bibnamefont
  {Sasaki}},\ }\bibfield  {title} {\bibinfo {title} {{Charmonium potential from
  full lattice QCD}},\ }\href {https://doi.org/10.1103/PhysRevD.85.091503}
  {\bibfield  {journal} {\bibinfo  {journal} {Phys. Rev. D}\ }\textbf {\bibinfo
  {volume} {85}},\ \bibinfo {pages} {091503} (\bibinfo {year} {2012})},\
  \Eprint {https://arxiv.org/abs/1110.0888} {arXiv:1110.0888 [hep-lat]}
  \BibitemShut {NoStop}%
\bibitem [{\citenamefont {Navas}\ \emph {et~al.}(2024)\citenamefont {Navas}
  \emph {et~al.}}]{ParticleDataGroup:2024cfk}%
  \BibitemOpen
  \bibfield  {author} {\bibinfo {author} {\bibfnamefont {S.}~\bibnamefont
  {Navas}} \emph {et~al.} (\bibinfo {collaboration} {Particle Data Group}),\
  }\bibfield  {title} {\bibinfo {title} {{Review of particle physics}},\ }\href
  {https://doi.org/10.1103/PhysRevD.110.030001} {\bibfield  {journal} {\bibinfo
   {journal} {Phys. Rev. D}\ }\textbf {\bibinfo {volume} {110}},\ \bibinfo
  {pages} {030001} (\bibinfo {year} {2024})}\BibitemShut {NoStop}%
\end{thebibliography}%

\end{document}